\documentclass{sadhana}
\usepackage{float}
\usepackage{placeins}
\usepackage{graphicx,amsmath,bm}
\usepackage{authblk}
\usepackage{tabularx}
\usepackage{booktabs}
\usepackage{diagbox}
\usepackage{url}
\usepackage{float}

\begin{document}
\title{Techno economic feasibility study of solar ORC in India}

\author{Ayona Biswas\textsuperscript{1,2}, Arindam Mandal\textsuperscript{3}, Aditya Bandopadhyay\textsuperscript{4}, Sourav Mitra\textsuperscript{4}\and Sandeep Saha\textsuperscript{3,*}}
\affilOne{\textsuperscript{1} Department of Mechanical Engineering, Indian Institute of Engineering Science and Technology, Shibpur\\}
\affilTwo{\textsuperscript{2} Department of Mechanical Engineering, Imperial College London\\}
\affilThree{\textsuperscript{3} Department of Aerospace Engineering, Indian Institute of Technology, Kharagpur\\}
\affilFour{\textsuperscript{4} Department of Mechanical Engineering, Indian Institute of Technology, Kharagpur}

\twocolumn[{
\maketitle

\begin{abstract}
Solar energy has enormous potential because there is a worldwide need to meet energy demands. Depleting non-renewable energy resources, increasing carbon emissions, and other environmental effects concern the scientific community to develop an alternative approach to electricity production. In this article, we present the study of a solar-powered Organic Rankine cycle considering Indian climatic conditions. Initially, we scrutinized seven working fluids and assessed their performance in the ORC at an evaporator pressure range of 9–30 bar and a mass flow rate range of 0.2 kg/s to 4.5 kg/s. For a fixed sink temperature of 298 K, we evaluate the system using four different power ratings of 2, 20, 50, and 100 kW based on four different source temperatures of 423 K, 403 K, 383 K, and 363 K. We estimate the system cost for each working fluid in each scenario separately. Our findings suggest that R 1233zd(E) is the optimum performing working fluid based on cost, cost-effectiveness, and environmental friendliness. We also notice that the estimated system scale cost is very competitive and could be a great alternative to the technologies already on the Indian market.

\end{abstract}

}]

\section{Introduction}

\begin{figure}[!t]
    \centering{
    \includegraphics[width=1\columnwidth]{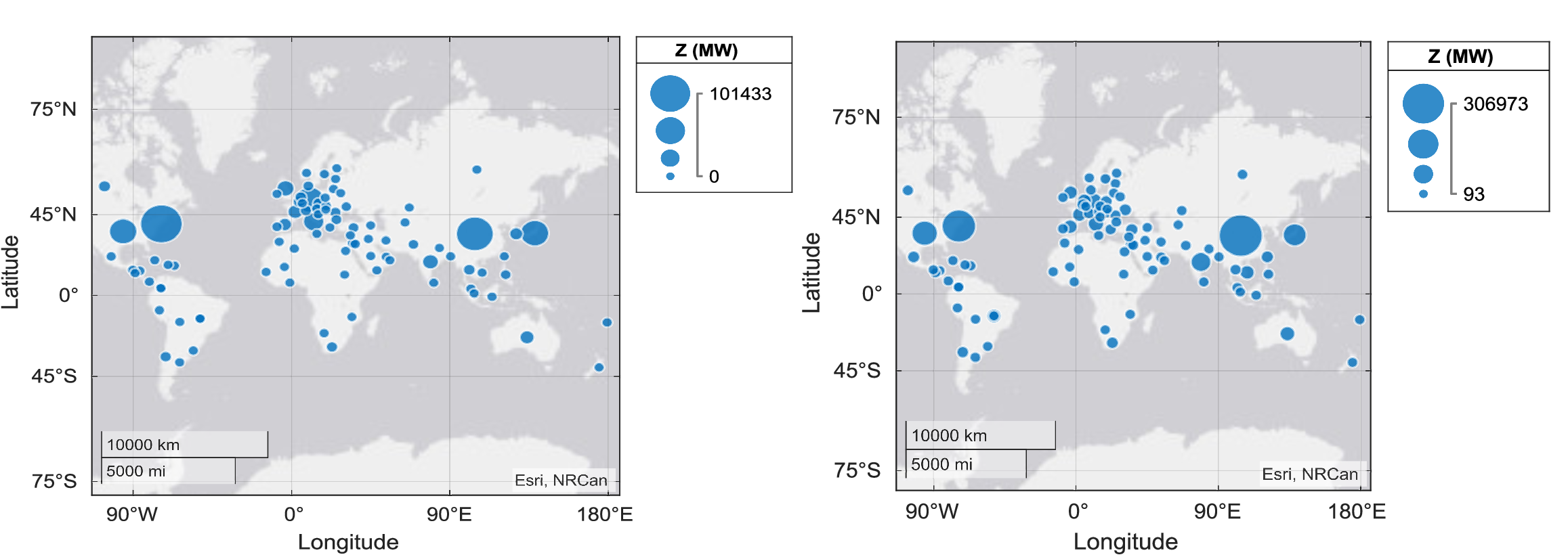}}
    \caption{Country-wise electricity production using solar energy \cite{WinNT1,WinNT2,WinNT3} for the year 2016 (left) and 2021 (right)}
    \label{solar_world_map}
\end{figure}

Conversion of renewable energy resources into electricity has been a top priority for almost all countries for several reasons. Existing challenges with conventional energy production, such as $CO_2$ emissions, depletion of non-renewable energy resources and other associated environmental impacts, suggest alternative clean energy production. Renewable resources like hydro, wind, biomass, and geothermal are strictly dependent on the geographic location and availability of the resource. Among the available renewable energy resources, solar energy holds the largest share due to its availability in almost all locations \cite{sharma2015jawaharlal}. A considerable number of nations currently use solar energy for electricity production. Figure \ref{solar_world_map} shows a substantial increment in solar-harnessed electricity production within five years. It is important to note that there still exists a considerable amount of disparity in solar power production. This increment is limited to a few countries as in USA, China, Germany, Japan, etc. Although reliance on renewable energy sources is now essential, many developing nations are still in the infancy stage due to their higher investment costs and lack of indigenous technology and infrastructure. For example, India emits more than 2.4 billion tonnes of carbon dioxide ($CO_2$) annually, which is considered "critically insufficient" to limit global warming to 1.5 degrees Celsius above pre-industrial levels. India should adopt renewable energy more swiftly than any other major country because it has the potential to build one of the largest solar energy farms in the world and because coal still accounts for more than 80\% of the nation's energy production. These restrictions provide us with a chance to investigate alternative solar energy harvesting technologies to meet both economic and environmental demands.

There are two primary ways of harnessing solar energy into electricity, (a) direct conversion of the energy for electricity production by employing solar PV cells and (b) by converting energy into heat and further use for electricity production. Despite the operational area of each of them differing according to the space requirement, solar thermal storage offers several advantages over PV systems in terms of investment cost, reliability, and simplicity \cite{patil2017techno}. For instance, solar PV modules in a solar farm have a more significant ecological impact compared to collector technology. In addition, the high temperature created by the solar panels reduces their effectiveness and installing a sophisticated cooling system to prevent that enhances the system complexity could significantly increase the cost \cite{ghadikolaei2021enviroeconomic}. Also, the increased toxicity and its corresponding environmental impacts \cite{kWak2020potential} make thermal energy storage devices an appealing alternative for solar power harvesting. The organic Rankine Cycle is a promising technology for converting heat into electricity which came into existence in 1980. This technology offers several advantages over traditional steam-based Rankine cycle technology at lower source temperatures and, thus, is considered to be a good candidate for low-power electricity generation (up to 100 $kW_e$ \cite{ashouri2017thermodynamic, tchanche2014heat, delgado2007double}). Investigating this technology in terms of its techno-economic feasibility can help in receiving acceptance in the global market, especially for developing countries, thus the present objective of this article. Though we analyze the potential of an ORC system in the purview of the Indian market, it will remain analogous for any specific country based on its market scenario.

Globally, S-ORC has a relatively long history. The first technical and economic feasibility study of a solar-powered ORC was investigated by Barber\cite{barber1978current} where the system achieved a maximum efficiency of 11\%. Since then, numerous efforts are present in the literature where the maximum thrust is given to the selection of organic fluids \cite{freeman2017working, gao2015working, habibi2020working, qiu2012selection, quoilin2012working, rayegan2011procedure, tchanche2009fluid, thurairaja2019working}, optimizing the cycle \cite{ayachi2014orc, wang2013thermodynamic, bellos2017parametric, boyaghchi2018assessment, delgado2010analysis, keshavarzzadeh2019multi, marion2012study, petrollese2019robust, quoilin2011performance}, assessing the system in component level\cite{declaye2013experimental, bellos2017parametric, loni2016optimizing, pantano2017expander, gao2015simulation, fiaschi2012thermo, qiu2011expanders, tiwari2020thermodynamic} techno-economic analysis \cite{patil2017techno, zare2015comparative, tocci2017small, quoilin2013techno, astolfi2015techno, pantaleo2020hybrid, keshavarzzadeh2019multi, heberle2017techno, aguilar2018techno}, etc. A number of application-based studies are reported in order to obtain a successful transition of the technology in domestic \cite{guarracino2016performance, freeman2017working, freeman2015assessment, ramos2018optimisation}, off-grid energy storage \cite{ozcan2017thermodynamic, petrollese2020ottana}, combined heat and power\cite{malavolta2010experimental, tempesti2012thermodynamic, tempesti2013thermo, calise2015design}, rural electrification \cite{seshie2018small, quoilin2013rural, orosz2009small}, tri-generation systems \cite{zhao2018solar, wu2019thermodynamic, shukla2019thermodynamic, sachdeva2019thermodynamic}, reverse osmosis \cite{bruno2008modelling, delgado2012design, nafey2010thermo, nafey2010combined, kosmadakis2009economic, delgado2007double, wegener20193e, zhao2018solar}, etc. Globally, there remain, several major technological challenges addressed in some literature \cite{ziviani2014advances, loni2021review, shahabuddin2021critical}. The sensitivity of the components (e.g., Collector, turbine, condenser) plays a crucial role in determining the effectiveness of the system. Solar collectors and the turbine have major contributions to the cost of the system \cite{chacartegui2016analysis, loni2021review, jebasingh2016review}. Cost-effectiveness and low efficiency of the system is somehow major challenge hindering the process of evolving this technology to the developing nations compared to the Western countries \cite{baral2015energy, quoilin2011performance, baral2018estimation}. 

In India and other developing countries, S-ORC dragged researchers' attention in late 2000. During the early days of development, the major focus was to design efficient solar collectors and Rankine-type heat engines for direct or indirect use \cite{gupta2003solar, reddy2012exergetic, desai2014simulation}. Several articles \cite{sarkar2015potential, desai2016thermo, sharma2021exergo, kshirsagar2015effect, dhalait2021selection, prajapati2020thermo} reviewed the thermo-economic analysis and systematic comparison of working fluids. Sahoo et al., Sircal et al. \cite{sahoo2015scope, sircar2017utilization} addressed the scope of solar-integrated biomass plants for the polygeneration process. While assessing a Udaipur-based plant, Bishoyi and Sudhakar \cite{bishoyi2017modeling, bishoyi2017modeling1} addressed some potential benefits of this technology using an LFR type of solar collector. The design of TES plays a crucial role in the performance of the system. Patil et al. \cite{patil2017techno} highlighted the reliability of an S-ORC over PVs based on five performance indicators. Several region-specific case studies have been reported by Wegener et al. \cite{wegener20193e}, Bist et al.\cite{bist2020heuristic}, Tripathy and Anand \cite{tripathi2021performance}, Bist et al. \cite{bist2020modeling}, Gogoi and Saikia \cite{gogoi2019performance}.

Current plans and initiatives of the government \cite{prajapati2019comparative, hairat2017100} indicate an increase in interest in this technology. Besides, several consortiums are presently active, involving more than 40 partners from industry, academia, and research labs in India and the United States, focusing on optimizing the cycle efficiency and reducing the collector cost \cite{ginley2018solar}.
Based on the data disclosed by the Ministry of New and Renewable Energy, Government of India, presently, India has achieved a massive 226\% increment in renewable energy installation and has a 25.64\% of share in total installed capacity \cite{hairat2017100, WinNT}. Furthermore, there is a plan to achieve 40\% of its stake in electricity production by 2030. Besides clear electricity production, solar energy can also be effective in district heating, clean hydrogen production, desalination, etc. As a result, the National Institute of Solar Energy, the Solar Energy Council of India, and several other industries are keen to use it in such applications.

However, with several preexisting challenges and scopes, S-ORC technology has significant research priority in India\cite{sarkar2015potential, sircar2017utilization, elavarasan2020comprehensive, sahoo2015scope, popat2018current, wang2021india}. Initially, our study identifies seven working fluids based on their properties and environmental impact. Furthermore, We conduct a parametric analysis indicating the power potential of the ORC at different evaporator pressures and mass flow rates. We consider the temperature in the evaporator at $10^\circ$ K superheated condition. We also analyze the collector area requirement for different power potentials based on the Indian average solar irradiance. Using the iso-power lines, we examine the system in four scenarios for a target power requirement of 2, 20, 50, and 100 kW. After that, we estimate the cost of the system starting from its component level for every working fluid. Finally, we identify the best working fluid candidate based on three parameters: Efficiency, system cost, and environmental friendliness. According to our assessment, S-ORC appears reasonably competitive in contrast to other existing renewable energy harvesting technologies.

The organization of this article are as follows: Section 2 describes the system and section \ref{criteria_for_ORC_fluid_selection} addresses the criteria adopted for the ORC fluid selection. Section \ref{consideration_for_cycle_design} discusses the assumptions considered for the analysis governing equation and section 3.3 describes the energy equations. In section \ref{results_and_discussion}, we report the solution methodology and a parametric study indicating the power potential of the ORC at different evaporator pressures and mass flow rates. Using the results we obtain in section \ref{results_and_discussion}, we evaluate the ORC in four separate scenarios at a target power of 2, 20, 50 and 100 kW in section \ref{scenario_based_evaluation}. In section \ref{system_scale_cost_estimation}, we perform a system scale cost estimation and identify the best working fluid suitable for the considered cycle. Finally, we compare the cost of ORC for the best-suited working fluid in section \ref{cost_comparison_v0p1} and summarize our key findings and the possible way forward in section \ref{conclusion}. 

\section{System description}

\begin{figure}[!t]
    \centering{
    \includegraphics[width=1\columnwidth]{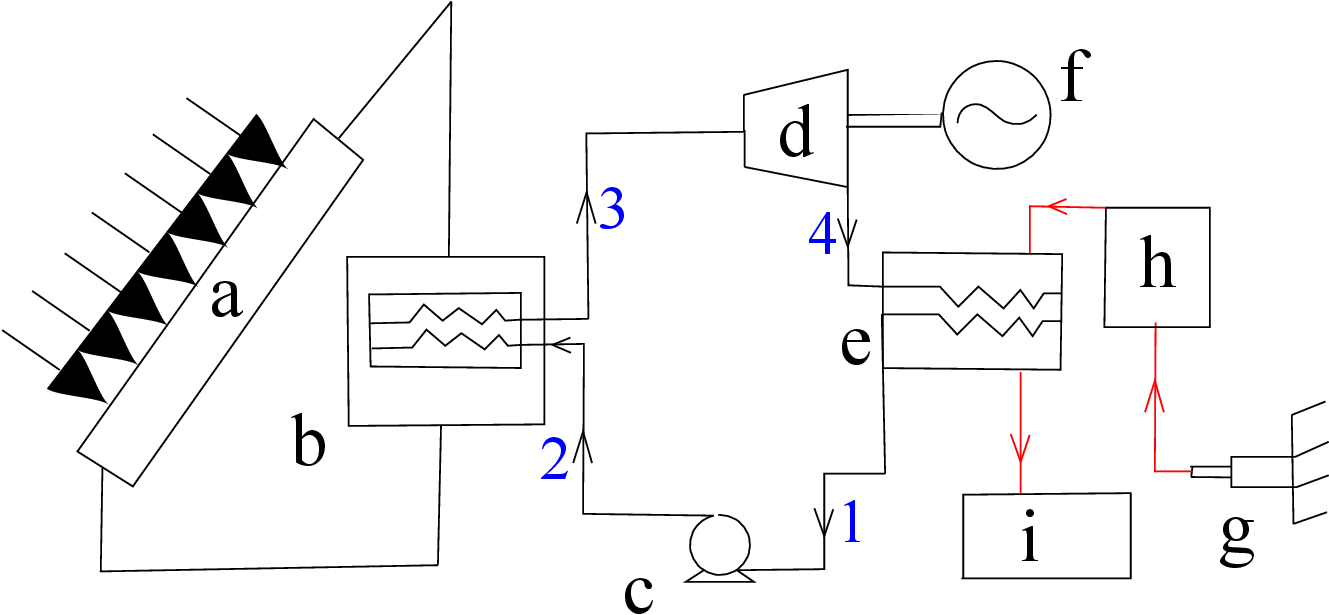}}
    \caption{Schematic diagram of the solar Organic Rankine Cycle. a. Solar collector, b. Evaporator integrated Thermal energy storage, c. Pump, d. Expander, e. Condenser, f.Generator, g. Water supply, h. Expansion vessel, i. Hot water tank.}
    \label{fig2}
\end{figure}

The S-ORC system we consider for the analysis is shown in figure \ref{fig2}. Black coloured stream is the working fluid circuit, whereas the red colour stream indicates a water stream that we consider a condensing medium. The components of the system are a solar collector, an evaporator integrated TES, a pump, an expander, a condenser, a generator, a water supply, an expansion vessel, and a hot water tank. We consider that the heat transfer fluid inside the thermal energy storage maintains a constant temperature during operation within a realistic range while using an evaporated tube collector. For the working fluid circuit, we consider four processes. In the liquid stage, the head of the working fluid is increased across the pump (between 1 and 2). Thereafter, heat addition takes place in the evaporator between steps 2 and 3, and the working fluid is converted into a high-pressure, superheated vapour. 4 to 5 indicates the expansion of superheated vapour in an expander, and the cycle is closed by the heat rejection process (4 to 1) inside a condenser. This work considers water at 298K as the condensing medium.

\section{Methodology}\label{methodology}
\subsection{Criteria for ORC Fluid selection}\label{criteria_for_ORC_fluid_selection}
To identify working fluids suitable for Indian climatic conditions, we consider the working fluids to have a relatively higher boiling point for mitigating the external effort required for condensation. In this work, We only consider the working fluids having A1, A2L, and B1 level safety standards as per ASHRAE classifications. Fluids with lower ODP and GWP are also preferred since they have less or no environmental impact. In addition, fluids with higher heat capacity, thermal conductivity, and vapour density are desirable for their higher heat-extracting capacity and lesser sizing requirements. Notably, The overall cycle performance, such as thermal efficiency, second law efficiency, and net power should be as high as possible, which is dependent on critical point, density, specific heat, acentric factor etc\cite{quoilin2011performance}. Finally, The cost and availability of ORC fluid in the Indian market are two important factors when developing a system with locally accessible resources. Fluids already used in the refrigeration or chemical industries are shortlisted and examined. The shortlisted fluids for the thermo-economic assessment are summarized in table \ref{shortlisted_fluids}. 

\begin{table}[htb]
    \centering
    \caption{Shortlisted working fluids and their properties \cite{lemmon2002nist}}
    \label{shortlisted_fluids}
    \scriptsize
    \renewcommand{\arraystretch}{0.65}
    \begin{tabularx}{\columnwidth}{ccccccccc}
    \hline
    Sl. & WF & T$_C$ & P$_C$ & NBP & AIT & ODP & GWP & Safety\\
    \hline
        1 & R 245fa & 426 & 36.1 & 288 & 685 & 0 & 858 &  B1 \\
        2 & R 11 & 470 & 43.7 & 302 & -- & 1 & 4660  & A1\\
        3 & R 601 & 469 & 33.6 & 309 & 533 & 0 & 4 & A3\\
        4 & R 601a & 460 & 33.7 & 301 & 693 & 0 & 4 &  A3\\
        5 & R 113 & 486 & 33.8 & 321 & 1043 & 1 & 5820 &  A1\\
        6 & R 123 & 256  &36.6  & 301 & 1043 & 0.02  & 79  & B1\\
        7 & R 1233zd(E)  &438.6 & 34.7 & 291 & 653 & 0 & 1 & A1\\
    \hline
    \end{tabularx}
\end{table}

\subsection{Consideration for cycle design}\label{consideration_for_cycle_design}
We consider four situations according to the temperature source, assuming that the HTF is at 423K for the first scenario and 403K, 383K, and 363K for the following three scenarios, respectively. The working fluid is further assumed to maintain a temperature having a $3^\circ$ pinch. We limit the evaporator pressure within its critical pressure and the turbine inlet temperature we consider to be at $10^\circ$ superheated condition. At a sink temperature of 298 K, the working fluid is recirculated to pump at a $5^\circ$ pinch. We fix the minimum line pressure in such a way that the fluid is in the liquid state at the condenser exit. In addition, we fix the efficiency of the pump and the expander to be 85\% and 70\%, respectively. Finally, we model the ORC considering the solar irradiance of 0.85 $kW/m^2$ and a collector efficiency of 75\%.

\begin{figure}[!t]
    \centering{
    \includegraphics[width=0.8\columnwidth]{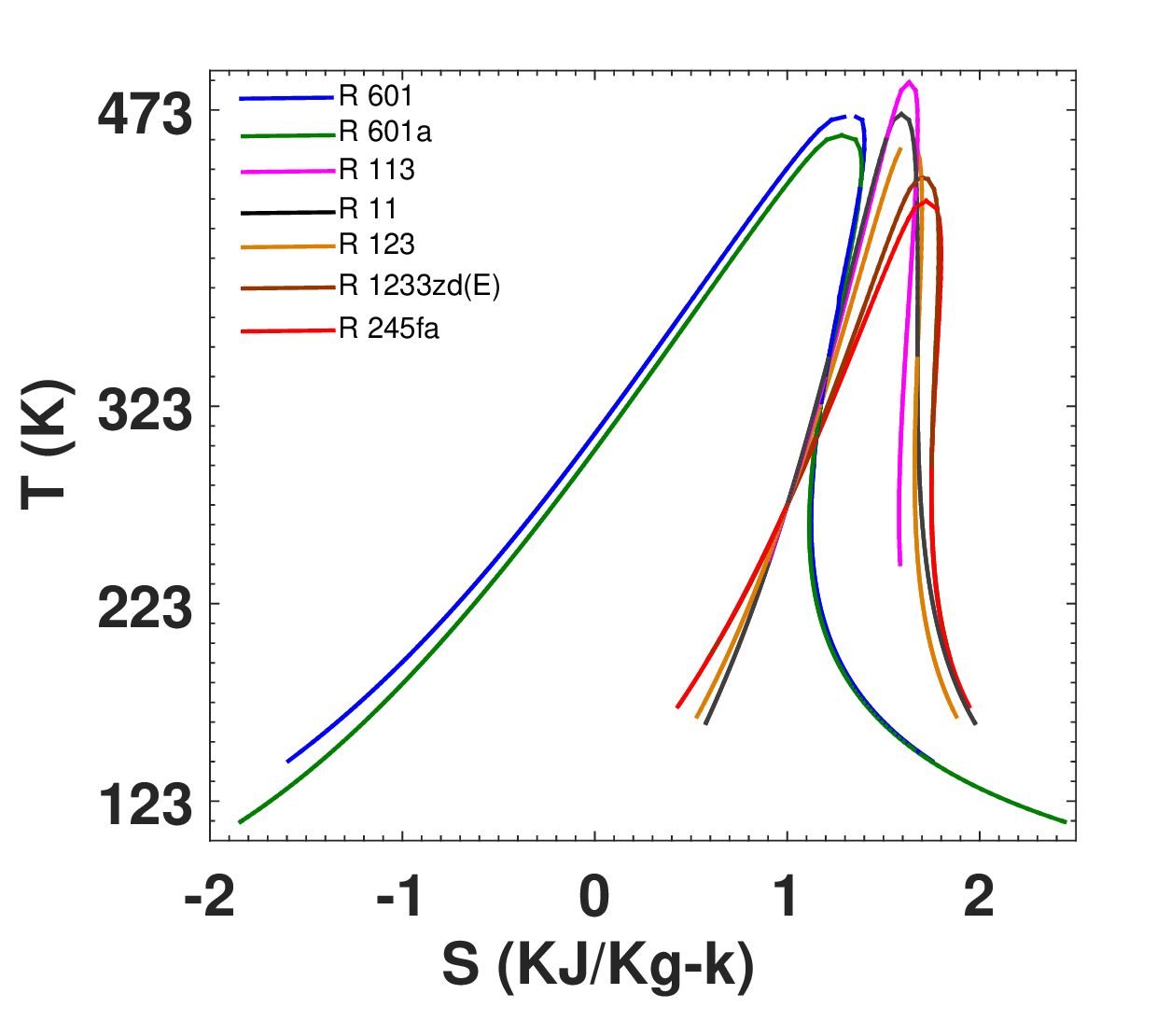}
    \caption{Temperature-entropy diagram of working fluids}
    \label{ts_diagram}}
\end{figure}

\subsection{Governing equations}\label{Governing_equation}

The governing equations for the components required for thermal analysis of the system are based on the following assumptions specified as follows; \emph (a) Steady-state steady flow, \emph (b) Negligible heat and pressure loss across the system, \emph (c)  Pump and Expander is working in reversible adiabatic condition. The energy balance equation can be written in the form of,
\begin{equation}\label{eq1}
\dot{Q}_{i} - \dot{W}_{o}+ \sum{\dot{m}_{i}h_{i}} - \sum{\dot{m}_{o}h_{o}} = 0
\end{equation}

\textbf{Pump modelling}: The power consumed by the working fluid pump is expressed as:  
\begin{equation}\label{eq2}
\dot{W}_{p}=\dot{m}(h_2-h_1)=\frac{\dot{m}v_1(P_1-P_2)}{\eta_{p}}    
\end{equation}

\textbf{Evaporator modelling}: The total heat transfer rate in the evaporator is expressed as: 
\begin{equation}\label{eq3}
\dot{H}_E=\dot{m}_f(h_3-h_2)= \dot{m}[(C_{P,f}(T_{E,in}-T_{E,out})+L]=\dot{Q}_{i}
\end{equation}

\textbf{Expander modelling}: The mechanical power is generated when the working fluid, in superheated vapour form, expands through the expander. The mechanical power generated by the expander is expressed as:
\begin{equation}\label{eq4}
\dot{W}_{t}=\dot{m}_f(h_3-h_{4s})\eta_{t}\eta_g= \dot{m}_f(h_3-h_4)\eta_g
\end{equation}

\textbf{Condenser Modelling}: The saturated vapour at the turbine outlet is subcooled to saturated liquid by rejecting heat to the cooling water in the condenser. The total heat rejection rate of the condenser is expressed as: 
\begin{equation}\label{eq5}
\dot{Q}_C=\dot{m}_f(h_4-h_1)= \dot{m}_lC_{P,l}(T_{C,out}-T_{C,in})
\end{equation}
\textbf{Solar collector Modelling}: The heat which is received by the solar collector and transferred to the heat transfer fluid can be calculated based on the collector energy balance expression as a function of collector efficiency:  \begin{equation}\label{eq6}
\dot{Q}_{i}= GA_{col}\eta_{col}= \dot{m}_{htf}C_{P,htf}(T_{col,out}-T_{col,in})
\end{equation}
The thermal efficiency of solar collector can be expressed in terms of solar irradiance:
\begin{equation}\label{eq7}
\eta_{col}=a_0-a_1\frac{T_{col,m}-T_{amb}}{G}-a_2\frac{(T_{col,m}-T_{amb})^2}{G}
\end{equation}
\textbf{Net power output and efficiency calculations}:\\
The net power output of the system is expressed as: 
\begin{equation}\label{eq8}
 \dot{W}_{Net}=\dot{W}_{t}-\dot{W}_{p}   
\end{equation}
The thermal efficiency of the system can be expressed as the ratio of net power output and heat input in the evaporator. 
\begin{equation}\label{eq9}
    \eta=\frac{\dot{W}_{Net}}{\dot{Q}_{i}}
\end{equation}

\section{Results and discussion}\label{results_and_discussion}
\subsection{Simulation methodology and parametric analysis}\label{simulation_methodology}

\begin{figure}[!t]
    \centering{
    \includegraphics[width=0.8\columnwidth]{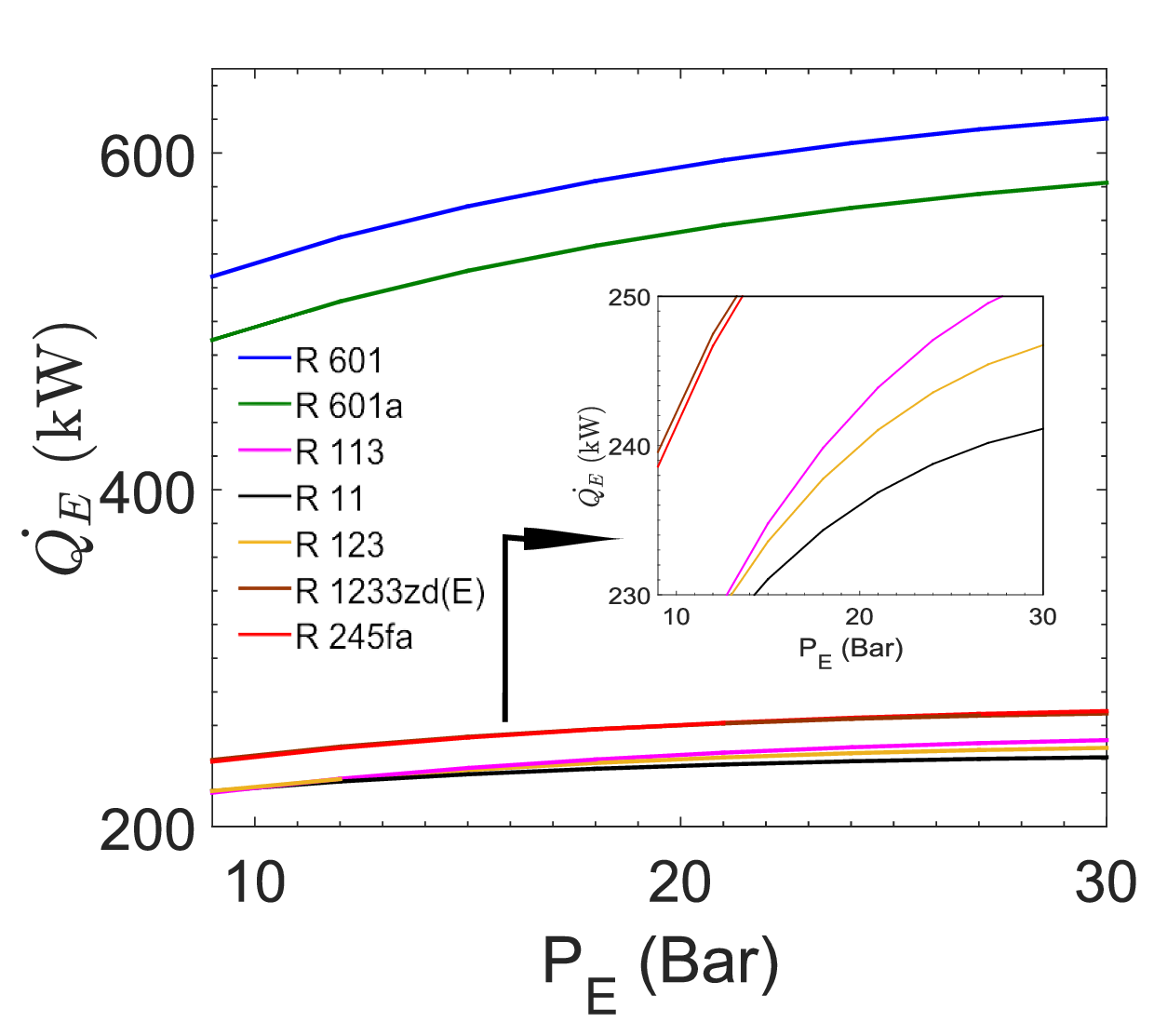}
    \caption{Heat addition rate in the evaporator w.r.t $P_E$}
    \label{w_in_all_fluids_inkskape}}
\end{figure}

\begin{figure}[!t]
    \centering{
    \includegraphics[width=0.8\columnwidth]{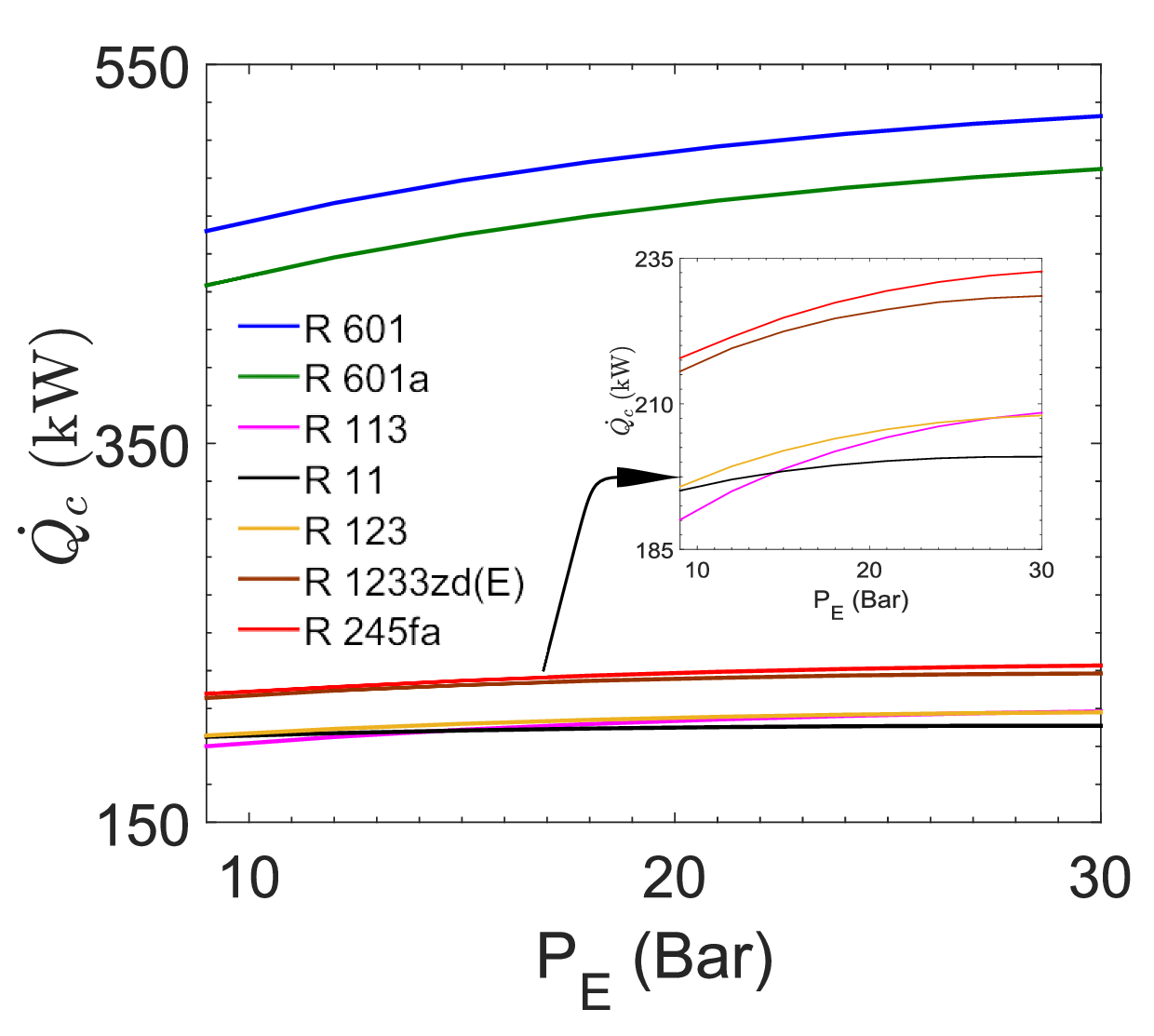}
    \caption{Heat rejection rate to the condenser at different $P_E$}
    \label{q_c_all_fluids_inkscape}}
\end{figure}

\begin{figure}[!t]
    \centering{
    \includegraphics[width=0.8\columnwidth]{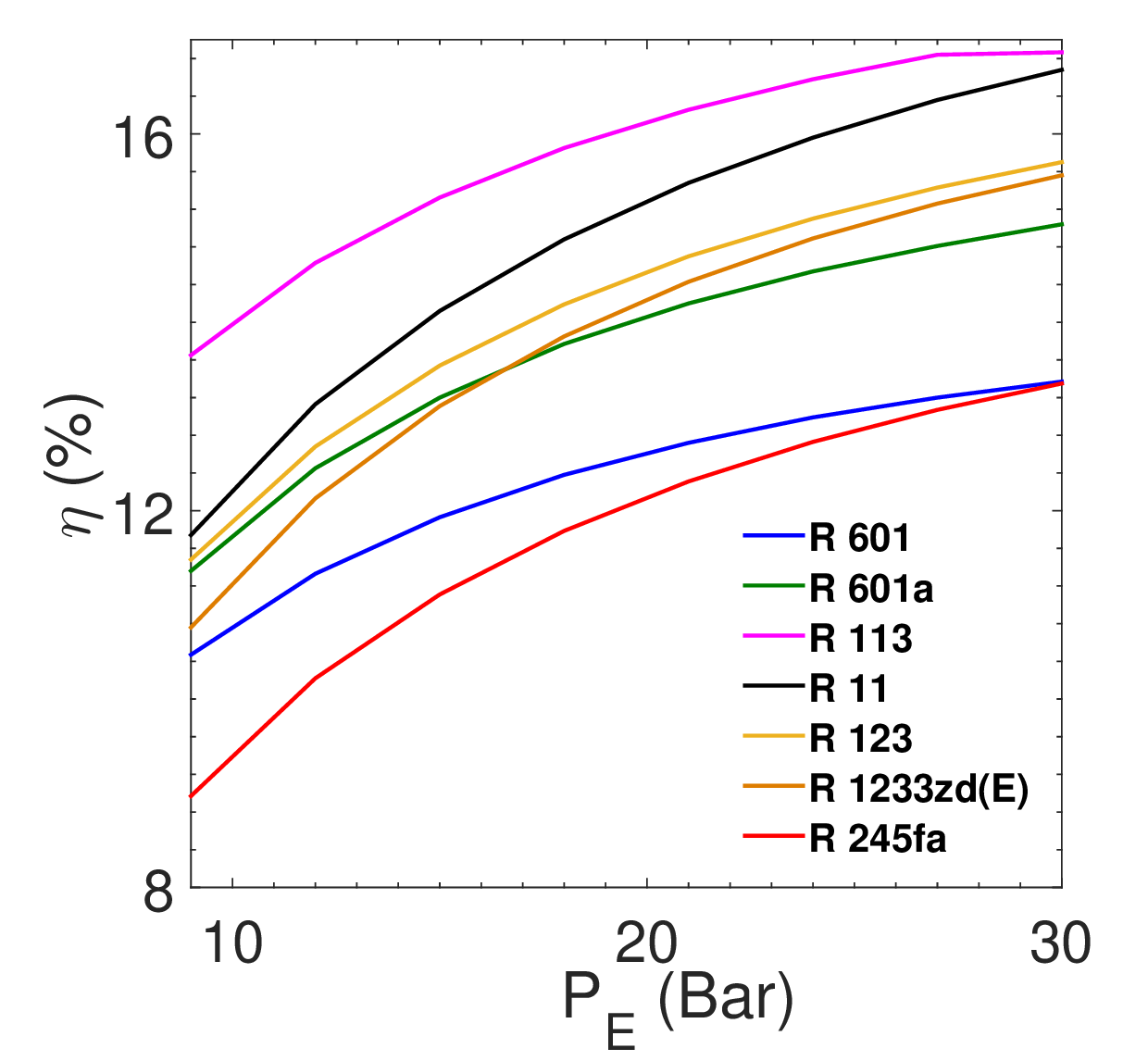}
    \caption{Cycle thermal efficiency at different $P_E$}
    \label{Eff_all_fluids}}
\end{figure}

\begin{figure}[!t]
    \centering
    \includegraphics[width=0.3\columnwidth]{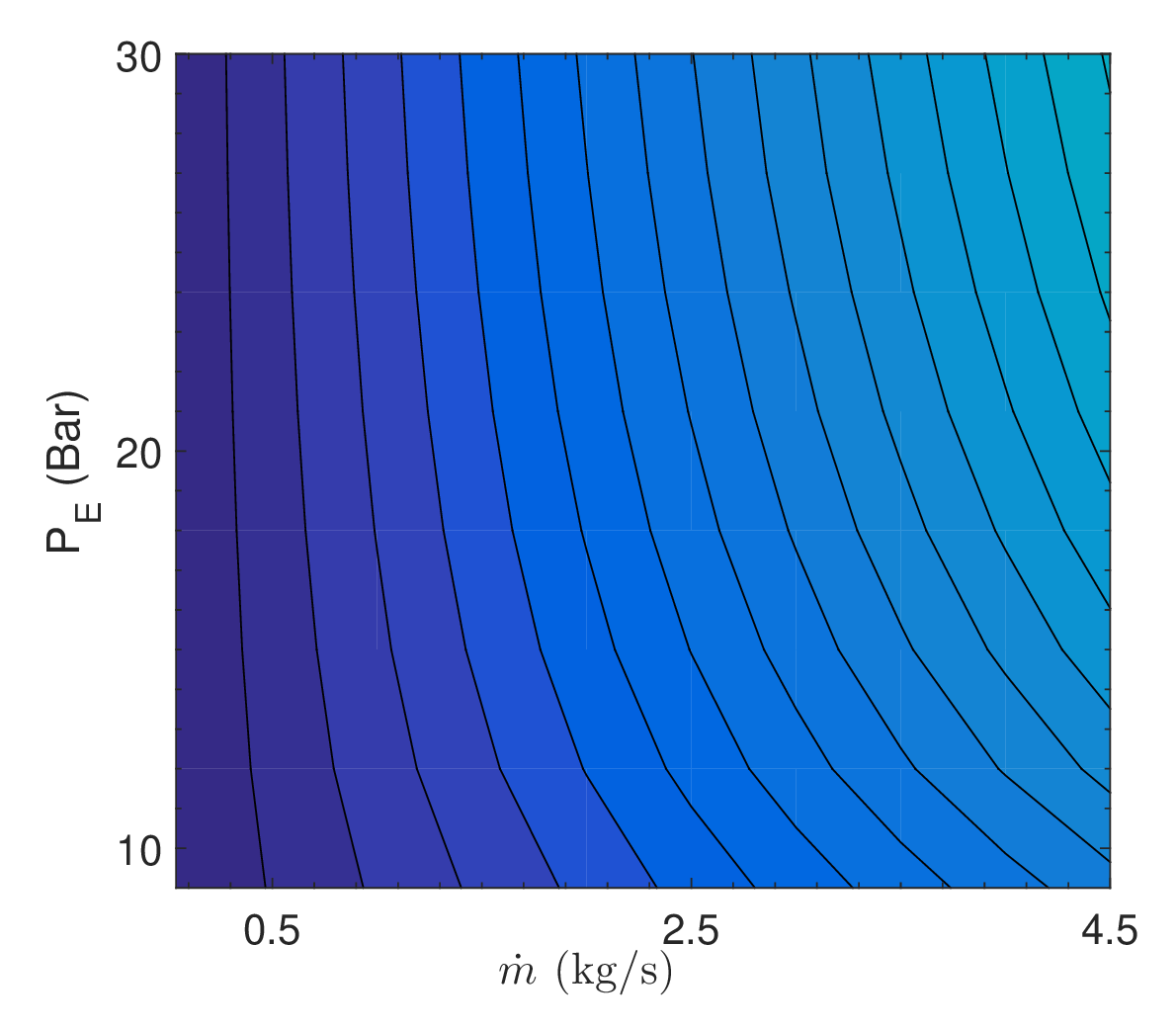}
    \includegraphics[width=0.3\columnwidth]{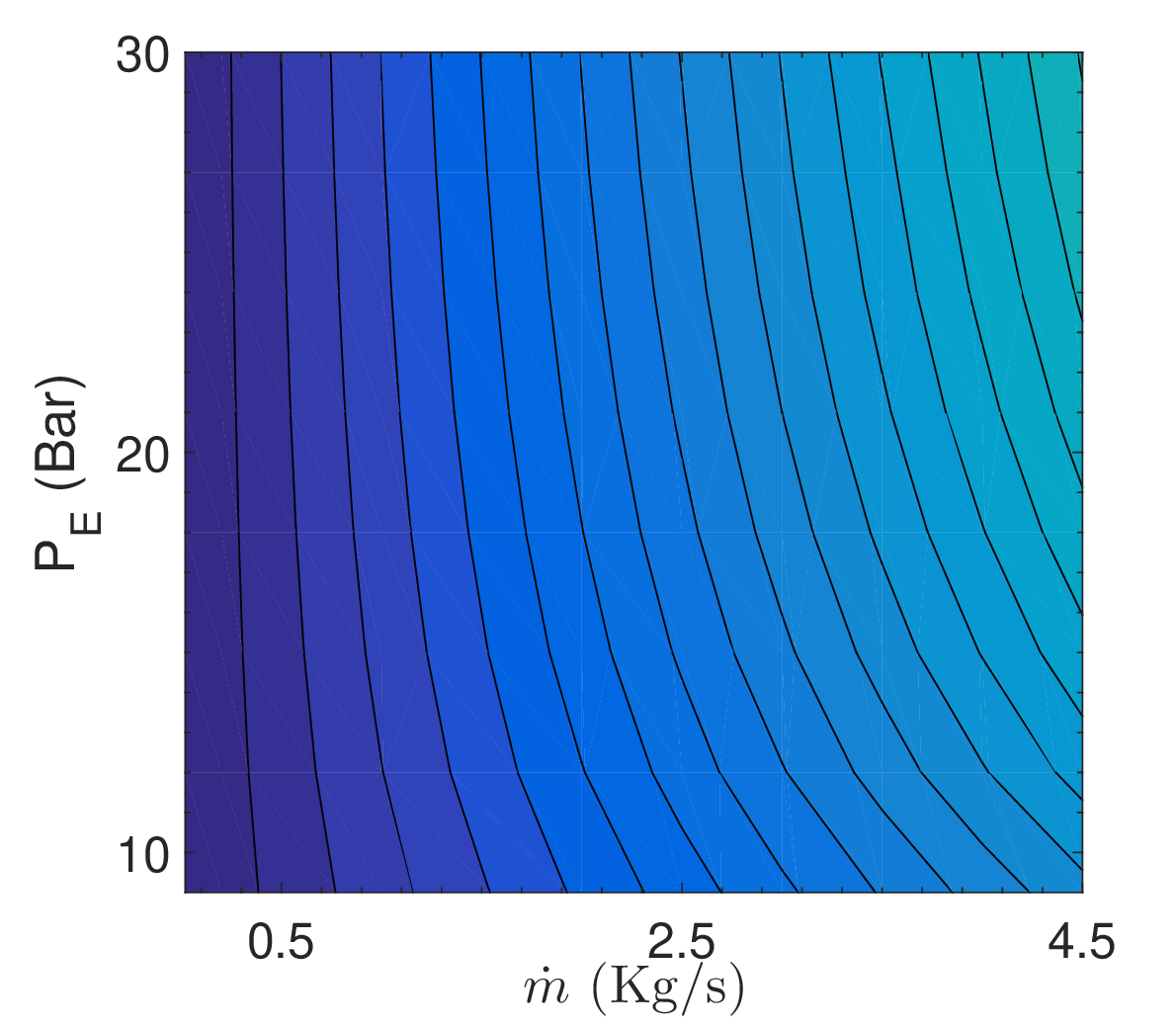}
    \includegraphics[width=0.3\columnwidth]{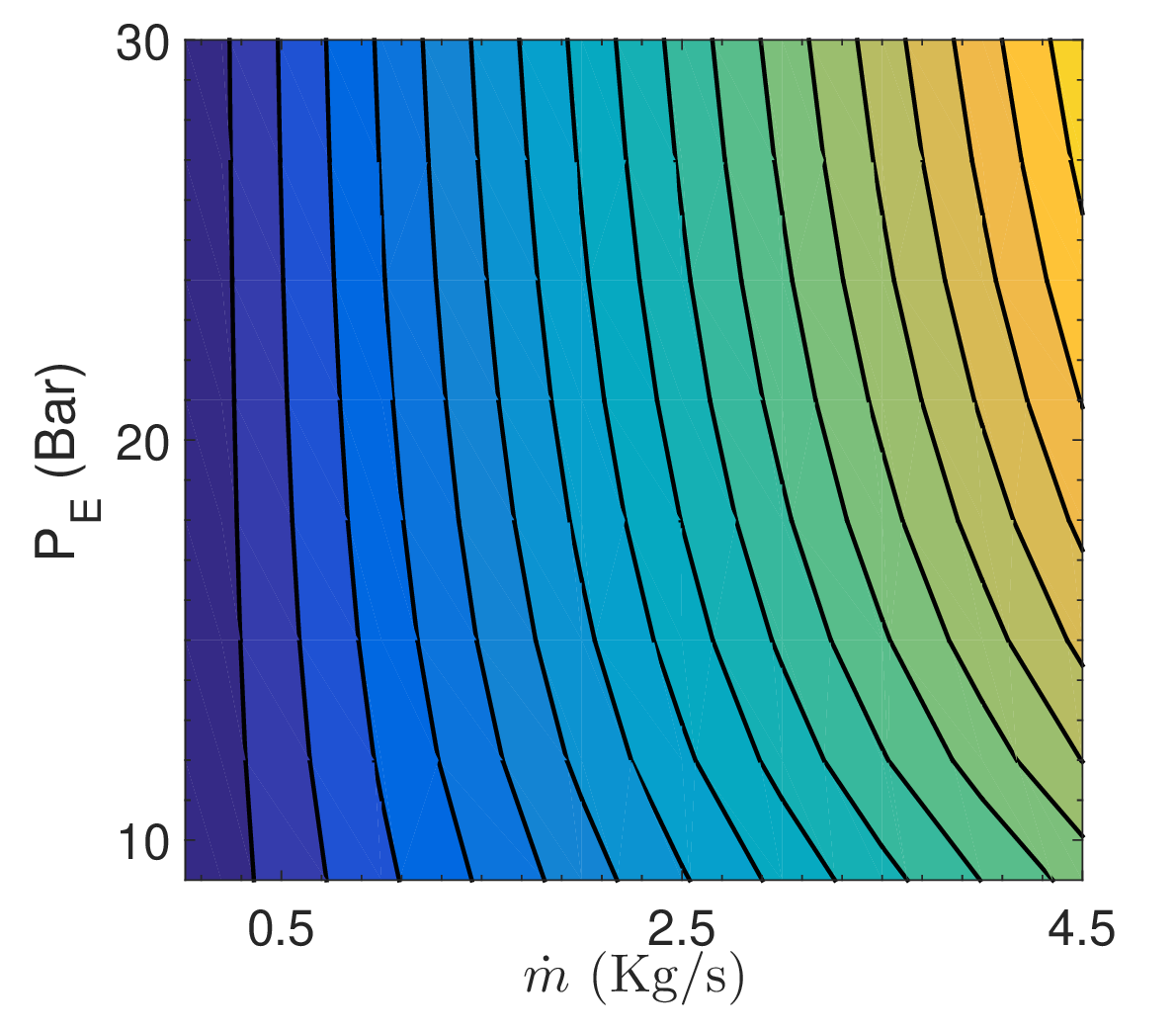}\\
    (a)\hspace{0.75in} (b) \hspace{0.85in} (c)\\
    \includegraphics[width=0.3\columnwidth]{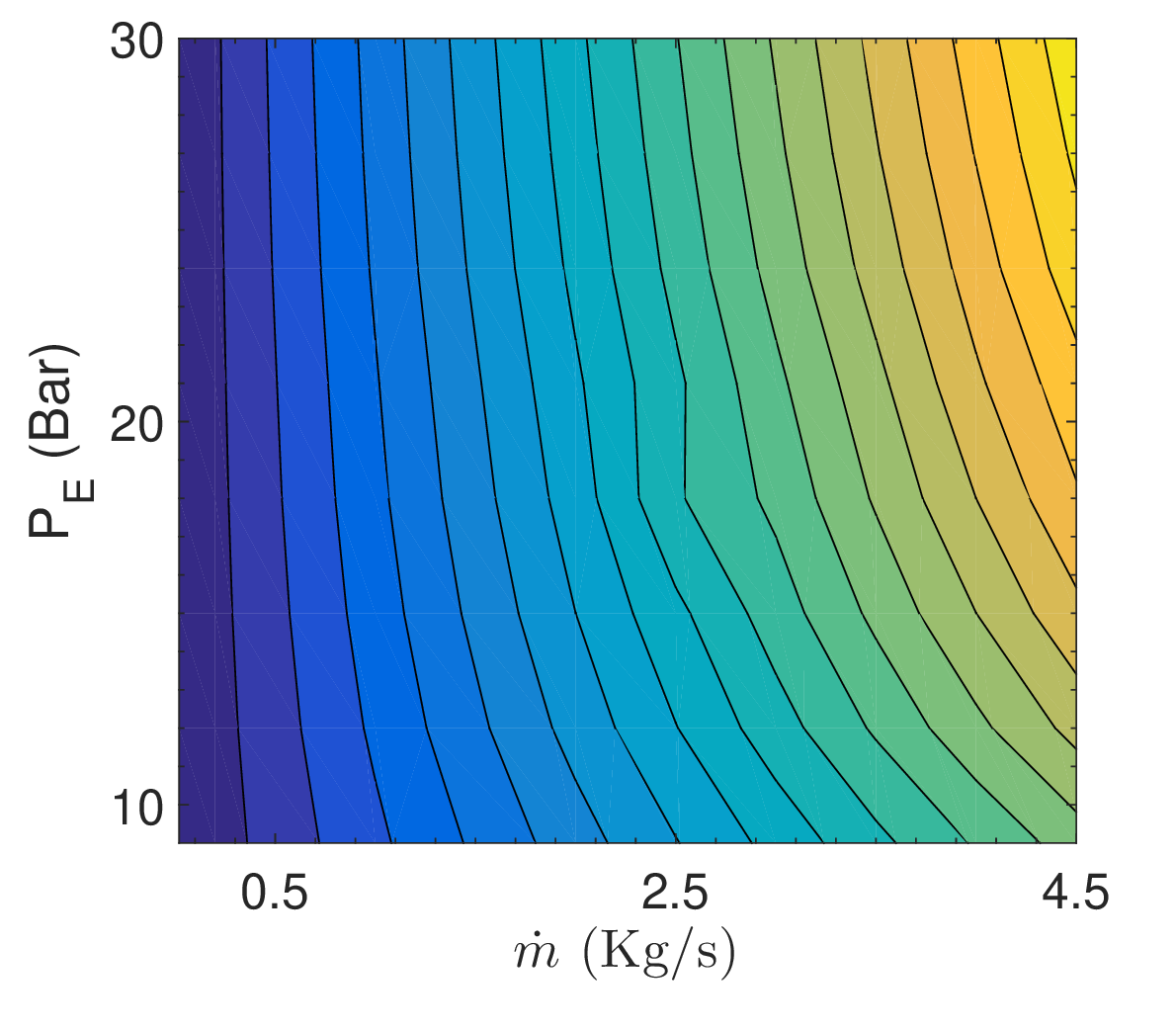}
    \includegraphics[width=0.3\columnwidth]{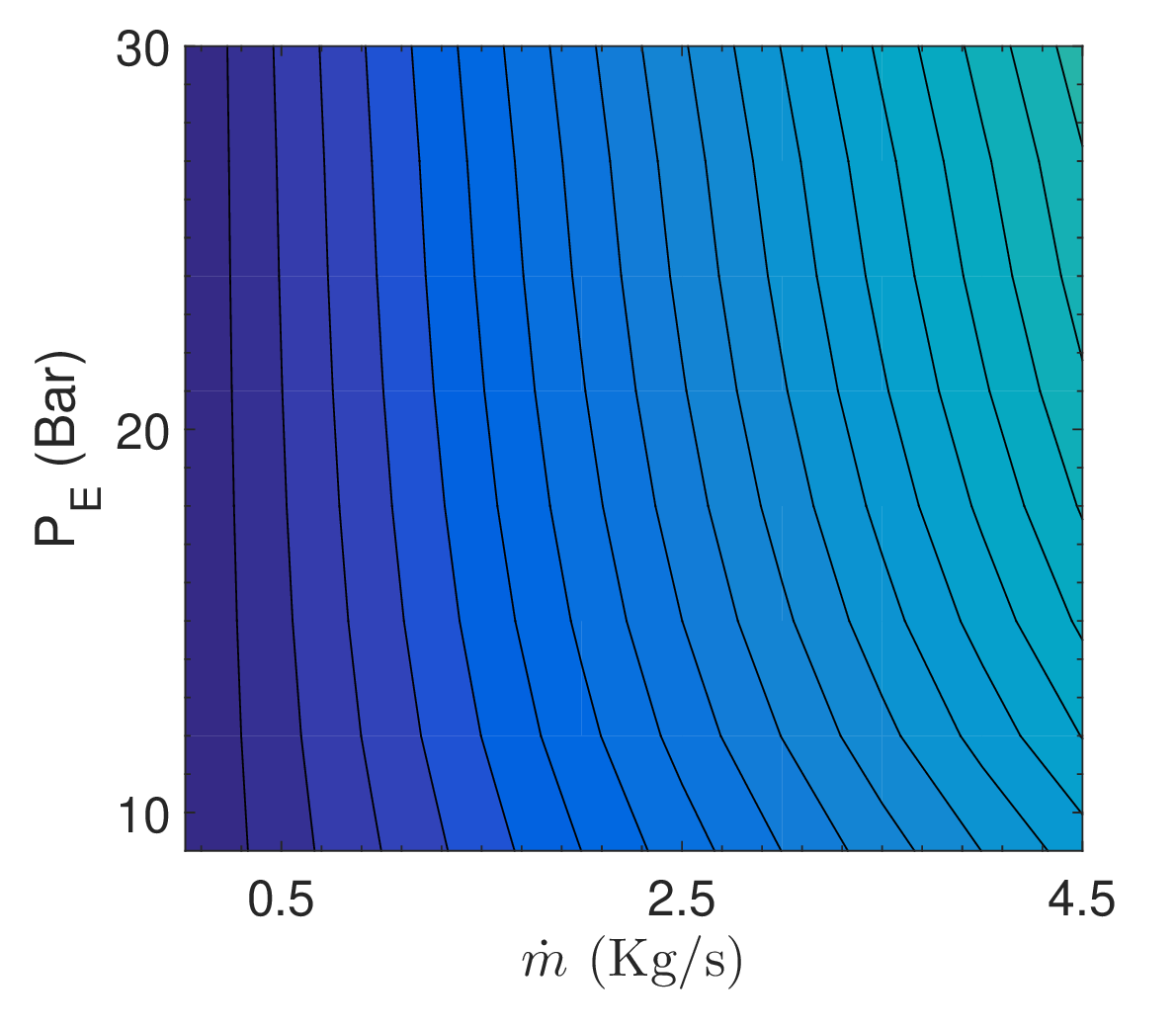}
    \includegraphics[width=0.3\columnwidth]{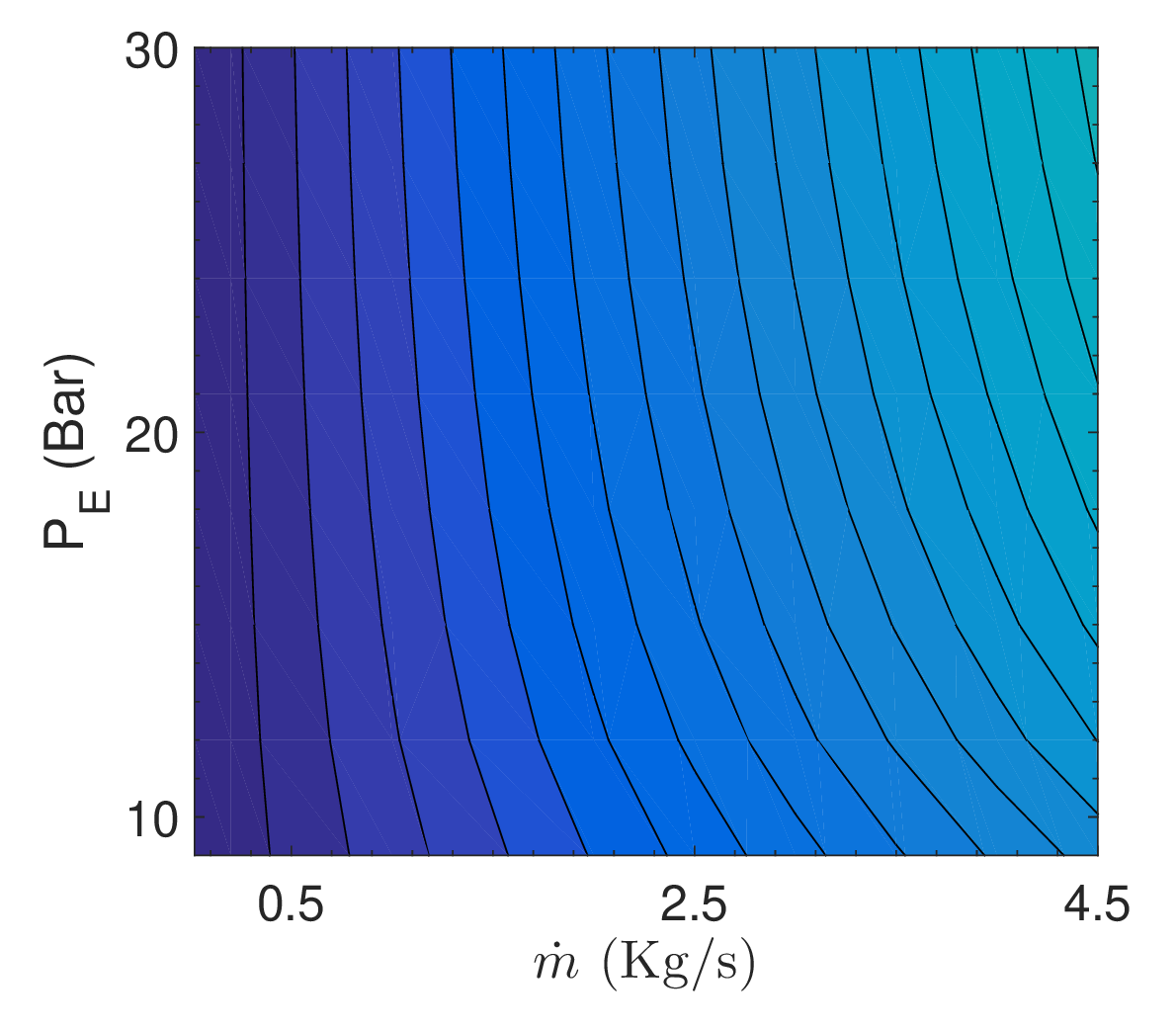}\\
    (d) \hspace{0.75in} (e) \hspace{0.85in} (f)\\
    \includegraphics[width=0.3\columnwidth]{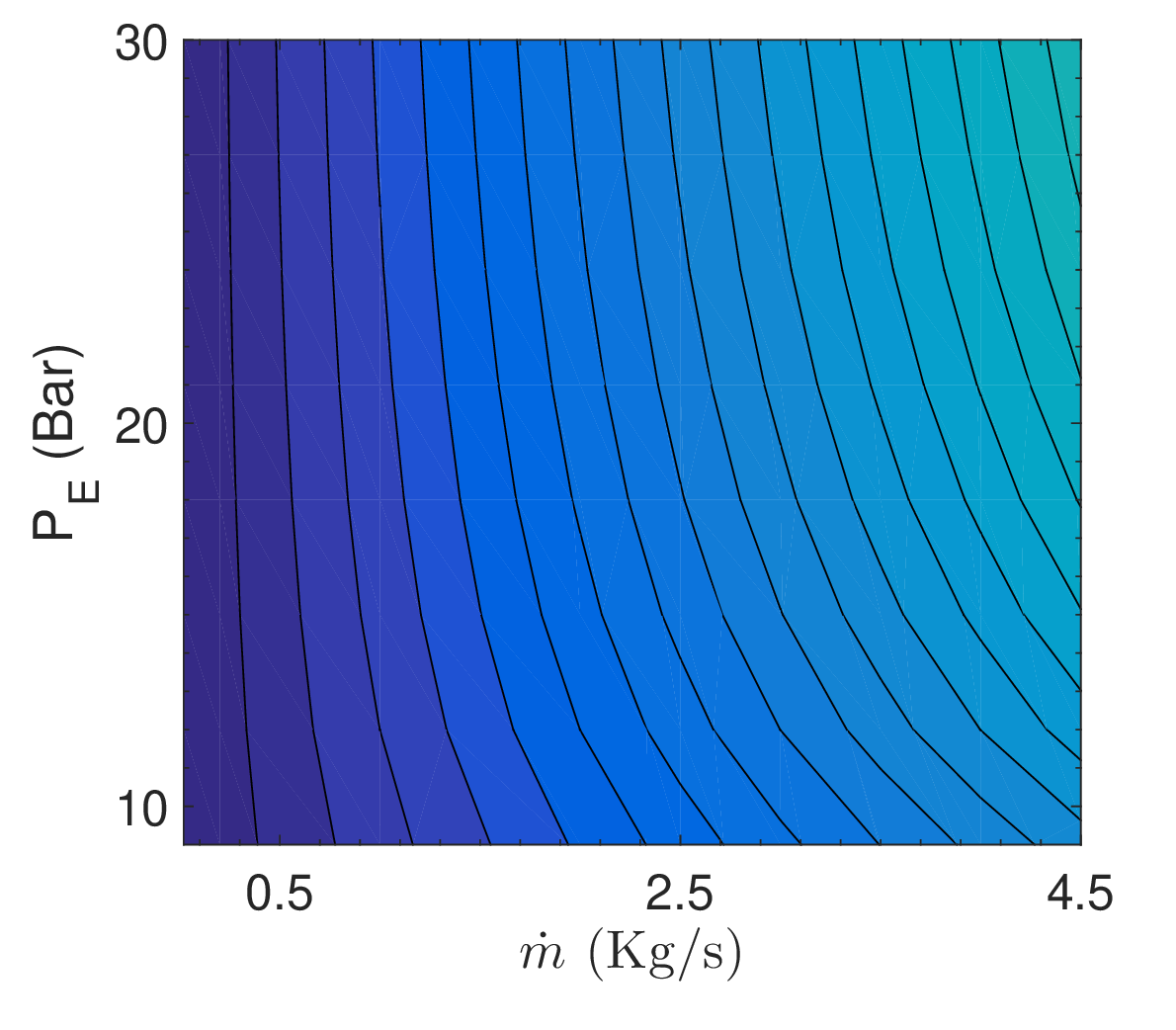}\\
    (g)\\
    \includegraphics[width=0.5\textwidth]{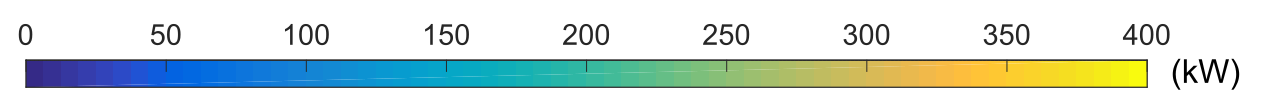}
    \caption{Contour of net power output for (a) R 245fa (b) R 11 (c) R 601 (d) R 601a (e) R 113 (f) R 123 (g) R 1233zd(E).}
    \label{isopower}
\end{figure}

After shortlisting seven organic working fluids shown in table \ref{shortlisted_fluids}, We conduct a series of numerical simulations for each fluid over the range of 9-30 bar of evaporator pressure and 0.2-4.5 kg/s of mass flow rate. Fluid properties are considered from the coolprop property database and the turbine inlet temperature is kept at $10^\circ$ superheated condition. The cycle is designed in DWSIM, an open-source process simulator tool. Furthermore, condenser pressure for each fluid is kept so that the fluid can achieve its liquid state at 303 K of ambient temperature. However, in the present analysis, we neglect the pressure drop across the heat exchanger. Figure \ref{isopower} indicates the power contour concerning the evaporator pressure ratio and mass flow rate for each shortlisted working fluid. R 601 and R 601a offer the maximum power output potential within the range of evaporator pressure and mass flow rate. The projected power output for R 601 and R 601a is almost twice that for the other fluids. As a result, the heat addition and rejection rate in the evaporator and condenser are also maximum for R 601 and R 601a (see figure \ref{w_in_all_fluids_inkskape} and figure \ref{q_c_all_fluids_inkscape}). For a fixed evaporator pressure, the heat addition and rejection rate for R 601 and R 601a are almost 2-3 fold higher than other fluids. Figure \ref{Eff_all_fluids} compares the cycle thermal efficiency for all the fluids where R 113 shows the highest and R 245fa offers the lowest trend in efficiency under the conditions. However, to make the comparison meaningful, we set the desired power output so that the available source and sink temperatures are assumed to be within the range of 363–423 K and 298 K, respectively.

\begin{figure}[!ht]
    \centering
    \includegraphics[width=0.3\columnwidth]{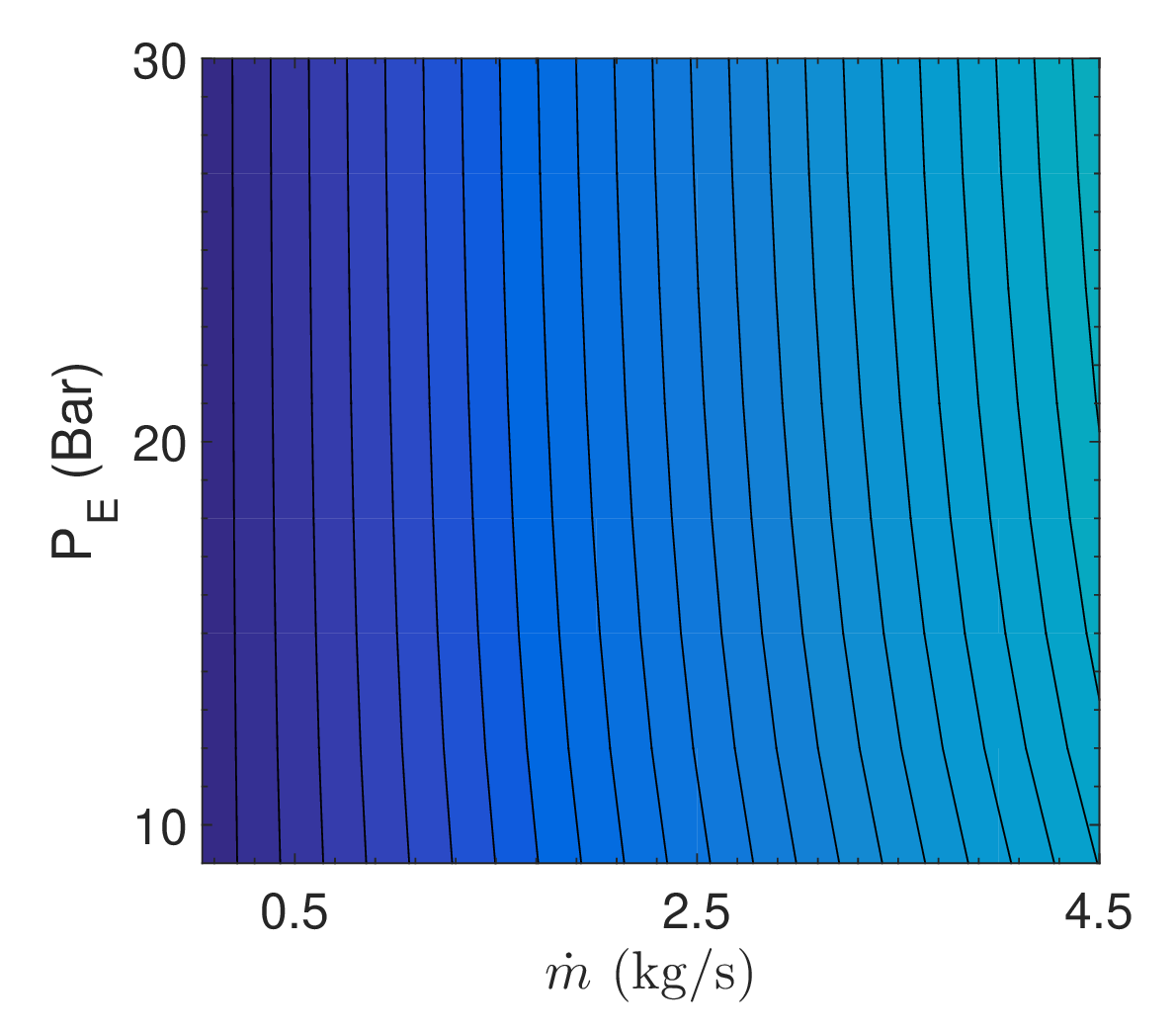}
    \includegraphics[width=0.3\columnwidth]{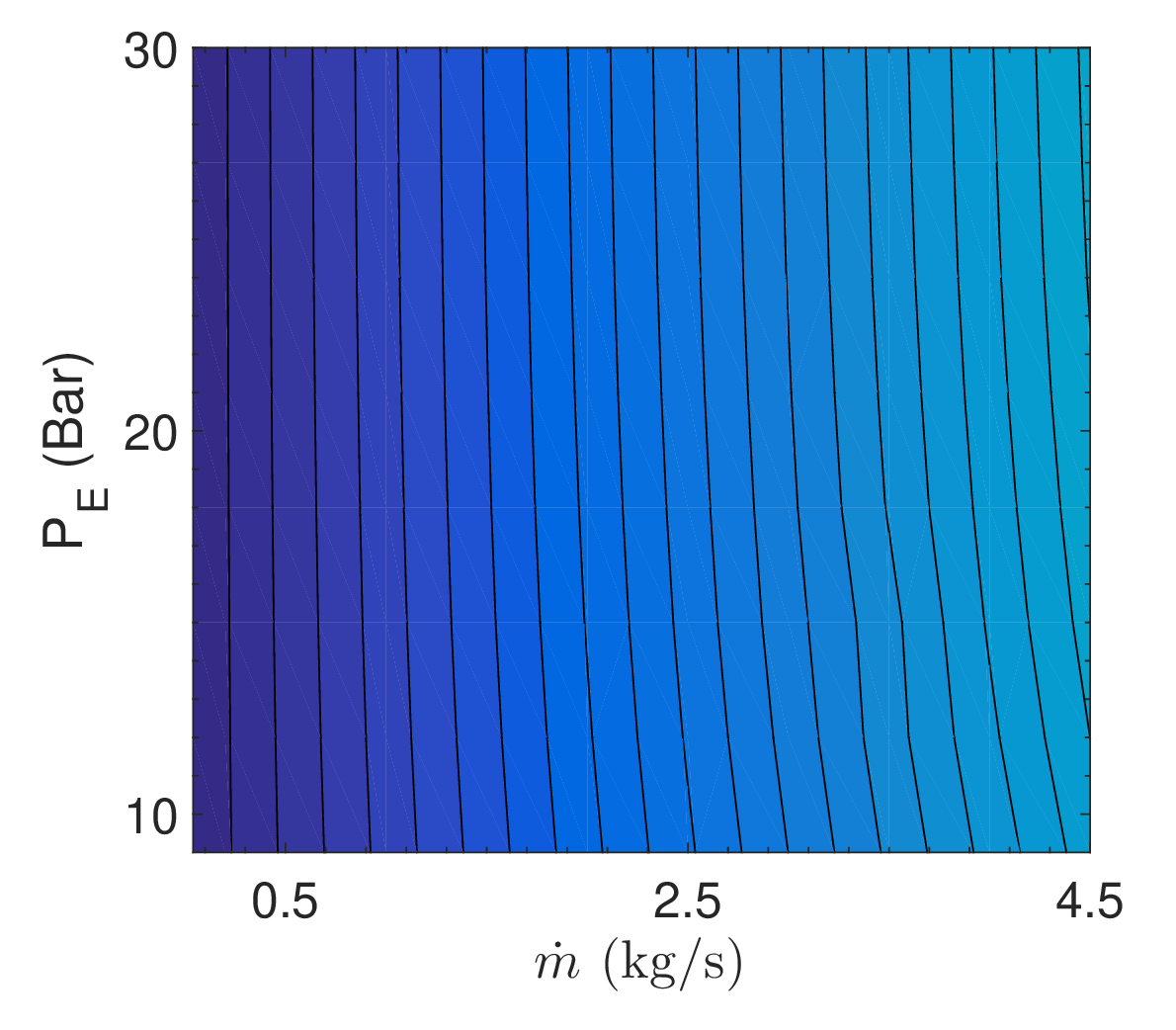}
    \includegraphics[width=0.3\columnwidth]{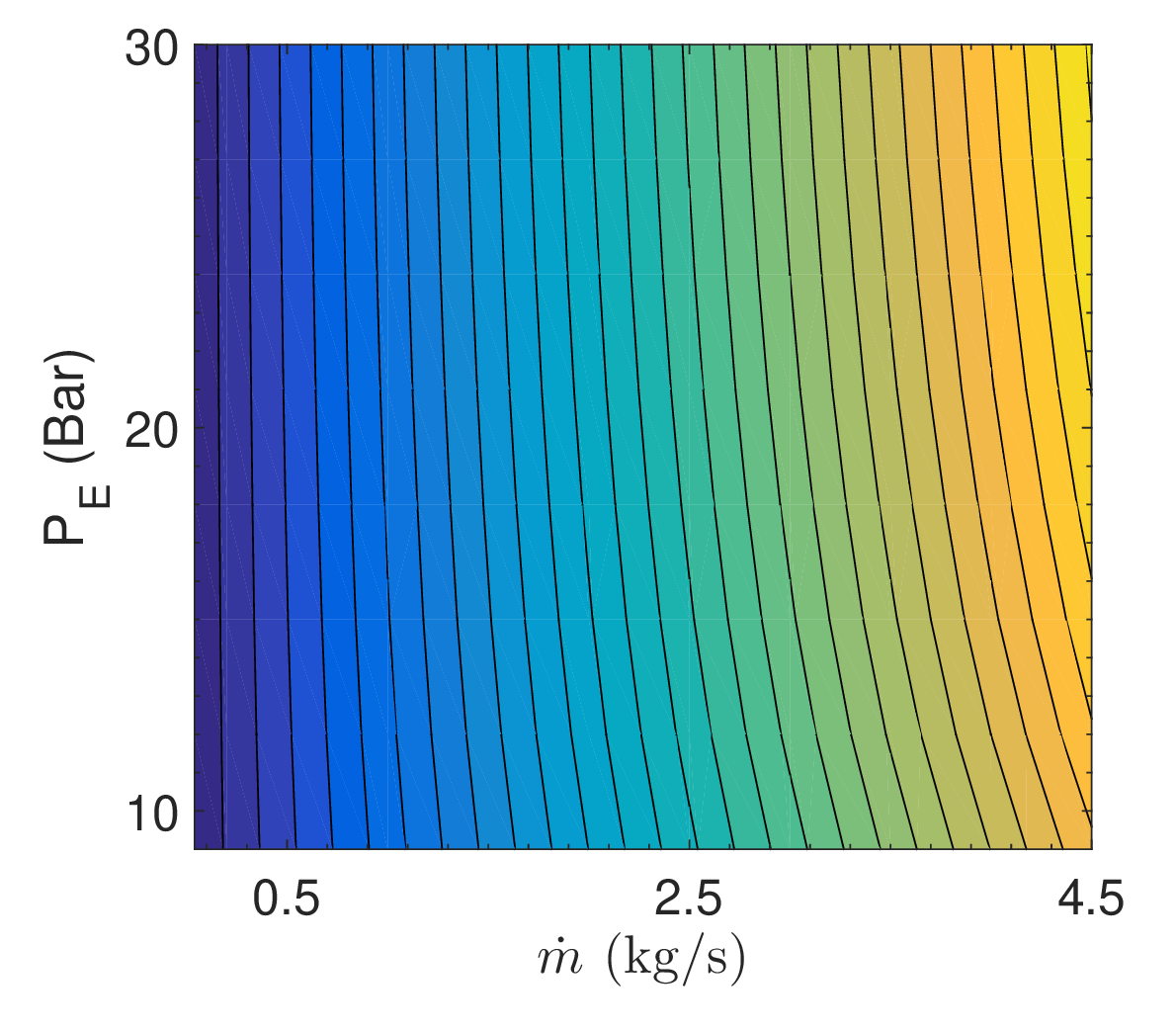}\\
    (a) \hspace{0.75in} (b) \hspace{0.85in} (c)\\
    \includegraphics[width=0.3\columnwidth]{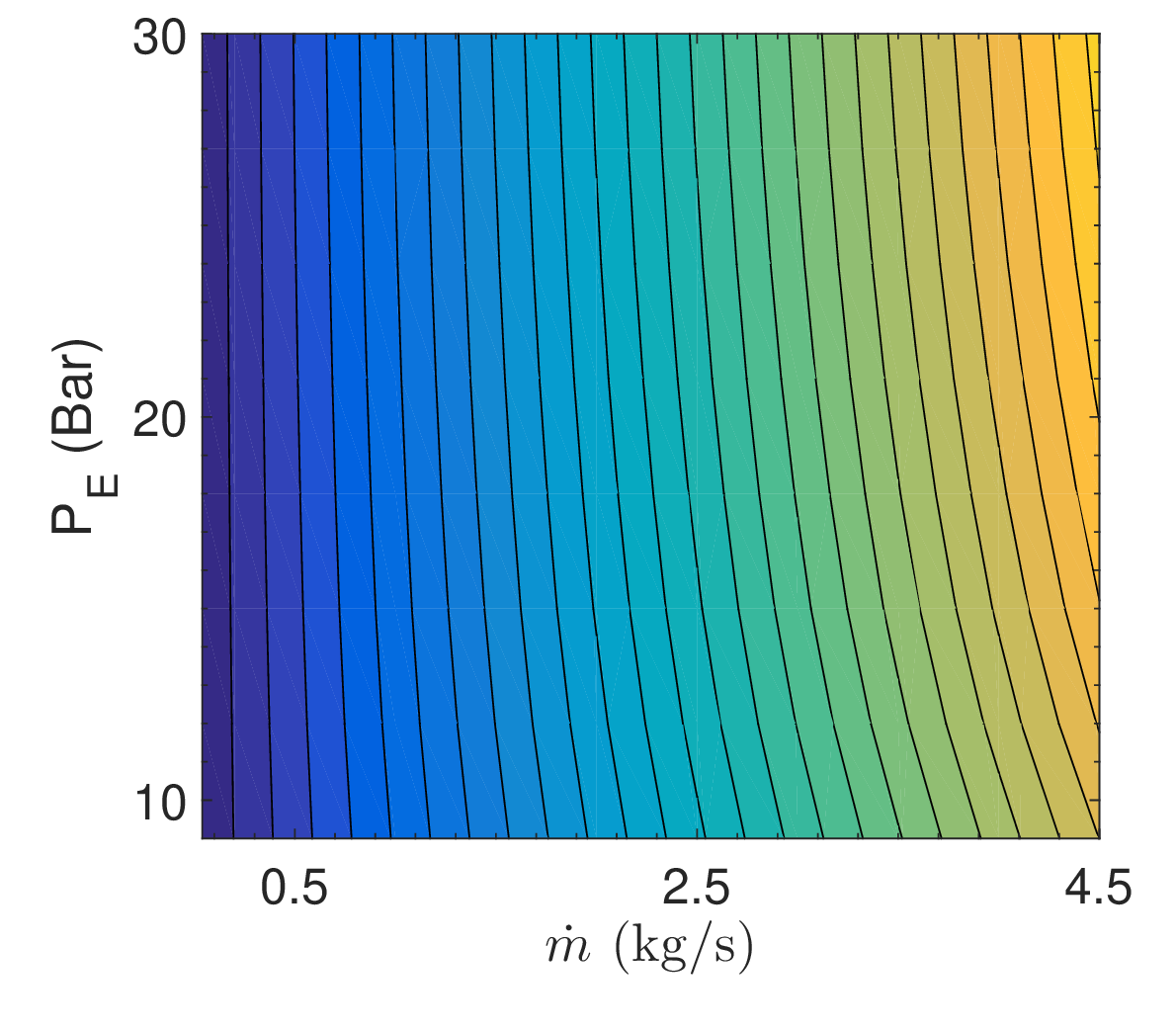}
    \includegraphics[width=0.3\columnwidth]{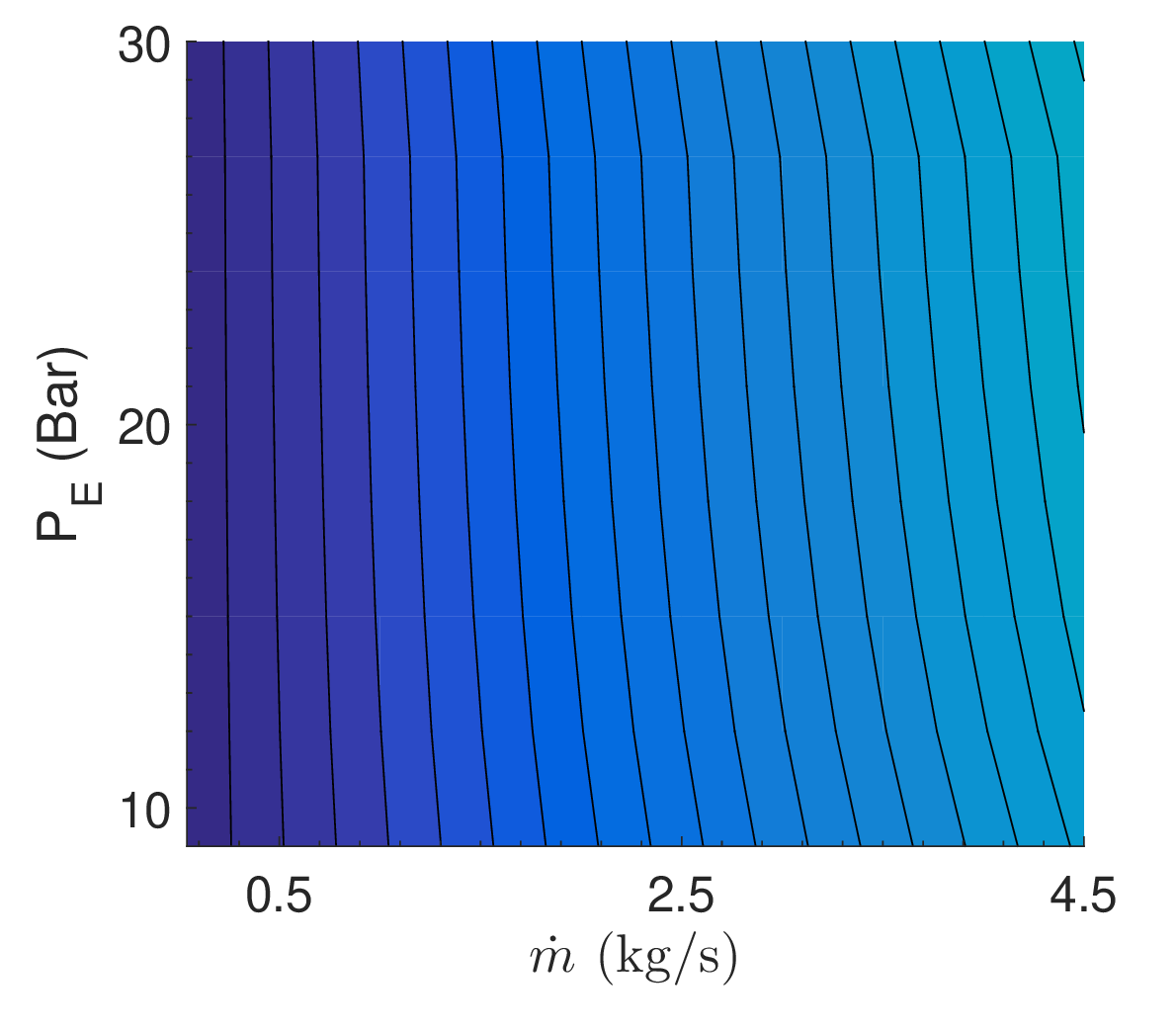}
    \includegraphics[width=0.3\columnwidth]{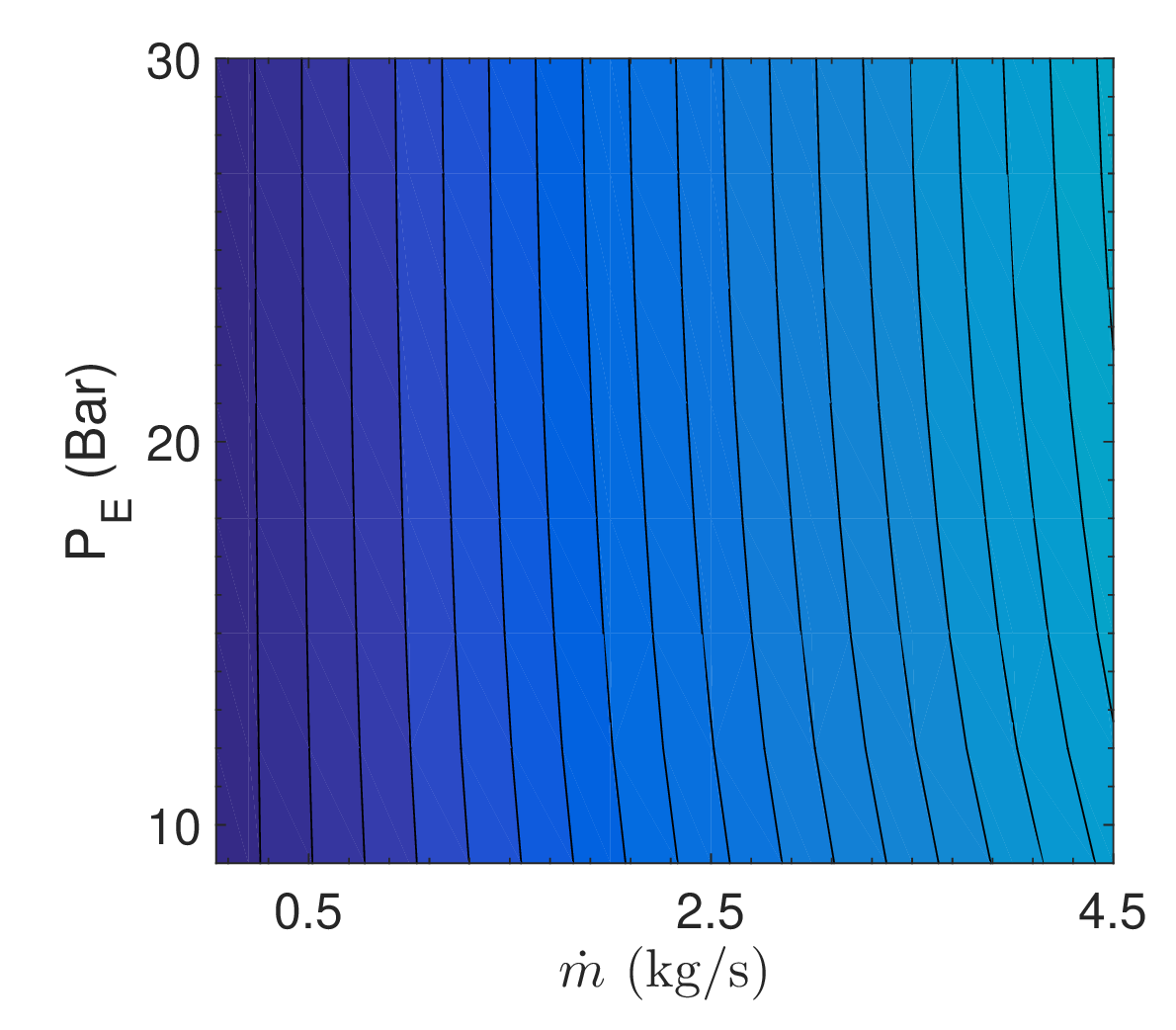}\\
    (d) \hspace{0.75in} (e) \hspace{0.85in} (f)\\
    \includegraphics[width=0.3\columnwidth]{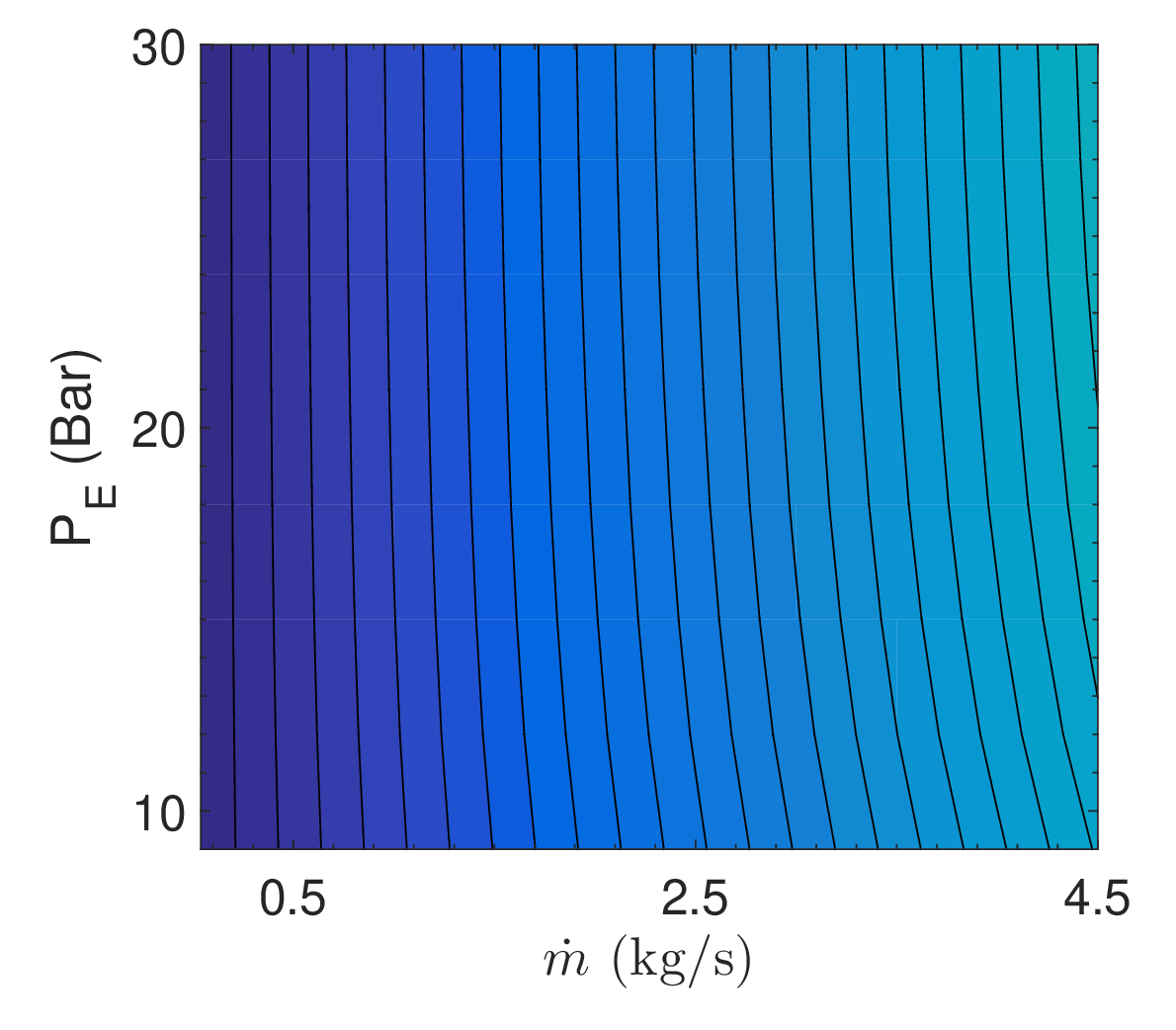}\\
    (g)\\
    \includegraphics[width=0.5\textwidth]{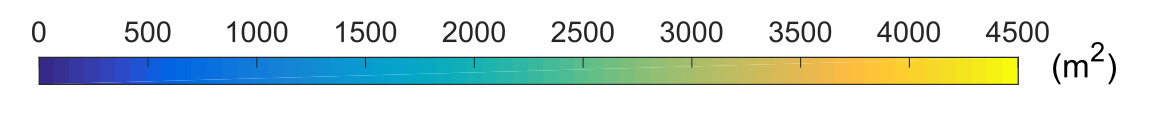}
    \caption{Contour of collector area requirement for (a) R 245fa (b) R 11 (c) R 601 (d) R 601a (e) R 113 (f) R 123 (g) R 1233zd(E).}
    \label{isoarea}
\end{figure}

\subsection{Operational envelope}\label{operational_envelope}
As per our assumption of $3^\circ$ and $5^\circ$ pinch at the evaporator and condenser, we fix the turbine inlet temperature of 420 K, 400 K, 380 K, and 360K for scenario 1 to scenario 4. Initially, we fix the pressure at the evaporator in such a way that the temperature is in $10^\circ$ superheated condition. The pressure at the turbine outlet is considered in such a way that the refrigerant becomes liquid at a temperature of 303 K. The properties of the selected fluids as described in figure \ref{ts_diagram} and table \ref{shortlisted_fluids} suggest different evaporator and line pressure requirements. We conduct four scenario-based evaluations of the target power requirement of 2, 20, 50, and 100 kW. The mass flow for each fluid is decided based on the isopower lines (see figure \ref{isopower}. For each case, we limit the operating temperature and pressure of the fluid within critical pressure and the source temperature. Notably, it is desirable to avoid wet expansion in the cycle for safe operation. Hence, this study does not include fluids having a negative temperature entropy slope. Presently out of the shortlisted fluids, only R-11 exhibits an infinite temperature entropy slope. Rest all the fluids have a positive slope of the temperature entropy curve.

\section{Scenario based evaluation}\label{scenario_based_evaluation}
In this section, We evaluate and compare the performance of each working fluid in four scenarios where, in each scenario, we keep the target power of 2, 20, 50, and 100 kW. The performance for each scenario is presented in table \ref{tab3} - table \ref{tab5}. However, we only present the result for a 2 kW ORC system. For higher power (e.g. 20 kW, 50 kW and 100 kW), $P_E$, $P_C$ $T_3$, $T_4$, $\eta$ would be the same whereas all other output parameters will be multiplied by 10 for 20 kW ORC, 25 for 50 kW ORC and 50 for 100 kW ORC system. The calculations for higher power output would further be used to estimate the system cost at the component level. The cycle efficiency lies within 13-15\% for scenario 1 and gradually reduces to 7-8\% for scenario 4 as the source temperature reduces from 423K to 363K. R 11, R 113 and R 1233zd(E) offer the highest cycle efficiency amongst all the fluids. For each case, we observe that N-pentane and R 113 require the lowest and highest flow rates, respectively. However, the minimum line pressure for n-pentane is lesser than the atmospheric pressure, which might cause adverse effects in the system. Notably, the operating expansion ratio for R 245fa is almost three-fold higher than R 113 for the same source temperature. It is interesting to note that, for scenario 1 (e.g. 423 K), fluid R 245fa requires the highest collector area whereas, for the other scenarios, R 123 requires the highest collector area. However, R 11 requires the lowest collecter area for all the scenarios. The output parameters shown in table \ref{tab2} - table \ref{tab5} play a significant role in determining the cost of the system.

\begin{table}[htb]
    \caption{Calculation table for scenario 1 ($T_{3}$ = 420 K)}
    \label{tab2}
    \centering
\scalebox{0.55}{\begin{tabular}{cccccccccccc}
    \hline
        WF & $P_{E}$ & $T_{4}$ & $P_{C}$ & $\eta$ & $A_{col}$ & $\dot{m}$ & $\dot{q_E}$ & $\dot{q_C}$ & $\dot{W_t}$ & $\dot{W_P}$ & $\dot{W_{net}}$\\
         \hline
        R-245fa	    & 26.681	& 348.84	& 1.77 & 13.04 	& 24.05 & .058 & 15.33 & 13.33 & 2.13 & 0.13 & 2 \\
        R-11	    & 16.873	& 335.49	& 1.25 & 14.62 & 21.22 & .058 & 13.53 & 11.55 & 2.05 & 0.07 & 2 \\
        R-601	    & 12.56	& 365.98	& 0.82 & 13.44 & 23.45 & .027 & 14.94 & 12.94 & 2.07 & 0.07 & 2 \\
        R-601a	    & 14.86	& 366.51	& 1.09 & 13.17 & 23.66 & .029 & 14.35 & 12.35 & 2.07 & 0.07 & 2 \\
        R-113	    & 9.55	& 362.65	& 0.54 & 13.86 & 22.62 & .065 & 14.42 & 12.42 & 2.04 & 0.04 & 2 \\
        R-123	    & 16.58	& 349.63	& 1.09 & 13.90 & 22.57 & .061 & 14.39 & 12.39 & 2.08 & 0.08 & 2 \\
        R-1233zd(E)	& 21.72	& 343.80	& 1.54	& 14.54 & 21.57 & .0525 & 13.76 & 11.75 & 2.10 & 0.10 & 2 \\
         \hline
\end{tabular}}
\end{table}

\begin{table}[htb]
    \caption{Calculation table for scenario 2 ($T_{3}$ = 400 K)}
    \label{tab3}
    \centering
\scalebox{0.55}{\begin{tabular}{cccccccccccc}
    \hline
        WF & $P_{E}$ & $T_{4}$ & $P_{C}$ & $\eta$ & $A_{col}$ & $\dot{m}$ & $\dot{q_E}$ & $\dot{q_C}$ & $\dot{W_t}$ & $\dot{W_P}$ & $\dot{W_{net}}$\\
         \hline
       R-245fa	    & 18.09	& 344.26	& 1.77 & 11.80 & 26.70 & 0.07 & 17.02 & 15.01 & 2.10 & 0.10 & 2 \\
       R-11	        & 11.63	& 331.78	& 1.25 & 12.98 & 24.12 & 0.068 & 15.38 & 13.38 & 2.06 & 0.06 & 2 \\
       R-601	    & 8.51	& 355.80	& 0.82 & 9.52 & 26.05 & 0.032 & 16.61 & 14.61 & 2.05 & 0.05 & 2 \\
       R-601a	    & 10.22	& 356.45	& 1.09 & 11.85 & 26.41 & 0.03 & 16.84 & 14.84 & 2.06 & 0.06 & 2 \\
       R-113	    & 6.37	& 353.32	& 0.54 & 12.35 & 25.42 & 0.08 & 16.20 & 14.20 & 2.04 & 0.04 & 2 \\
       R-123	    & 11.25	& 348.98	& 1.09 & 10.60 & 29.60 & 0.08 & 18.87 & 16.87 & 2.07 & 0.07 & 2 \\
       R-1233zd(E)	& 14.81	& 342.08	& 1.54 & 12.14 & 25.81 & 0.07 & 16.46 & 14.46 & 2.08 & 0.08 & 2 \\
         \hline
         \end{tabular}}
\end{table}

\begin{table}[htb]
    \caption{Calculation table for scenario 3 ($T_{3}$ = 380 K)}
    \label{tab4}
    \centering
\scalebox{0.55}{\begin{tabular}{cccccccccccc}
    \hline
        WF & $P_{E}$ & $T_{4}$ & $P_{C}$ & $\eta$ & $A_{col}$ & $\dot{m}$ & $\dot{q_E}$ & $\dot{q_C}$ & $\dot{W_t}$ & $\dot{W_P}$ & $\dot{W_{net}}$\\
         \hline
       R-245fa	    & 11.78	& 337.80	& 1.77 & 10.15 & 30.89 & 0.08 & 19.70 & 17.70 & 2.07 & 0.07 & 2 \\
       R-11	        & 7.70	& 327.34	& 1.25 & 10.95 & 28.72 & 0.84 & 18.30 & 16.30 & 2.05 & 0.05 & 2 \\
       R-601	    & 5.5183	& 345.40	& 0.82 & 10.29 & 30.50 & 0.04 & 19.44 & 17.44 & 2.04 & 0.04 & 2 \\
       R-601a	    & 6.74	& 345.02	& 1.09 & 10.15 & 30.96 & 0.04 & 19.73 & 17.73 & 2.05 & 0.05 & 2 \\
       R-113	    & 4.0521	& 343.70	& 0.54 & 10.49 & 29.89 & 0.10 & 19.06 & 17.06 & 2.02 & 0.02 & 2 \\
       R-123	    & 7.32	& 340.91	& 1.09 & 9.01 & 34.84 & 0.10 & 22.21 & 20.21 & 2.05 & 0.05 & 2 \\
       R-1233zd(E)	& 9.73	& 335.83	& 1.54 & 10.37 & 30.24 & 0.08 & 19.28 & 17.28 & 2.06 & 0.06 & 2 \\
         \hline
         \end{tabular}}
\end{table}

\begin{table}[htb]
    \caption{Calculation table for scenario 4 ($T_{3}$ = 360 K)}
    \label{tabcost}
    \label{tab5}
    \centering
\scalebox{0.55}{\begin{tabular}{cccccccccccc}
    \hline
        WF & $P_{E}$ & $T_{4}$ & $P_{C}$ & $\eta$ & $A_{col}$ & $\dot{m}$ & $\dot{q_E}$ & $\dot{q_C}$ & $\dot{W_t}$ & $\dot{W_P}$ & $\dot{W_{net}}$\\
         \hline
       R-245fa	    & 7.29	& 330.44	& 1.77 & 7.98 & 39.32 & 0.11 & 25.06 & 23.06 & 2.05 & 0.05 & 2 \\
       R-11	        & 4.85	& 322.64	& 1.25 & 8.46 & 37.18 & 0.11 & 23.70 & 21.70 & 2.03 & 0.03 & 2 \\
       R-601	    & 3.40	& 335.11	& 0.82 & 8.05 & 38.86 & 0.05 & 24.78 & 22.79 & 2.03 & 0.03 & 2 \\
       R-601a	    & 4.24	& 335.59	& 1.09 & 7.97 & 39.42 & 0.06 & 25.13 & 23.13 & 2.04 & 0.04 & 2 \\
       R-113	    & 2.43	& 334.06	& 0.54 & 8.16 & 38.40 & 0.13 & 24.48 & 22.48 & 2.02 & 0.02 & 2 \\
       R-123	    & 7.32	& 332.43	& 1.09 & 7.02 & 44.59 & 0.14 & 28.43 & 26.43 & 2.04 & 0.04 & 2 \\
       R-1233zd(E)	& 6.09	& 328.94	& 1.54 & 8.11 & 38.70 & 0.11 & 24.68 & 22.68 & 2.05 & 0.05 & 2 \\
         \hline
         \end{tabular}}
\end{table}

\section{system scale cost estimation}\label{system_scale_cost_estimation}
The cost of a system depends on several factors e.g., primary and ancillary components, flow rate, line pressure, power, installation, instrumentation, and others. The rudimentary cost estimates we provide in this paper are based on pre-existing literature that is connected to the current Indian market environment. We estimate the system scale cost of a solar-powered ORC in this part and evaluate its specific investment cost using the information we gather from our analysis based on the four scenarios. We divide the cost estimations into four sections. 

\begin{enumerate}
    \item Cost of Main ORC components includes the Evaporator, expander, condenser, and pump.
    \item Area-related solar thermal cost including Solar collector and Heat transfer fluid.
    \item Cost of the working fluid.
    \item Area-related solar thermal cost such as solar collector cost, HTF cost.
    \item Other cost includes Thermal energy storage, Hot water tank, Expansion vessel, and Instrumentation cost. 
\end{enumerate}
However, we limit the cost estimation to the component level for scenario 1 only. The specific investment cost for each scenario is detailed in table \ref{tabcost}.

The cost for the expander can be estimated using the correlation described by Dumont et al. \cite{dumont2019comparison},

\begin{equation}\label{eq15}
Cost_{E} = 250 \times \dot{W_t}
\end{equation}

The cost for heat exchangers (condenser or evaporator) can be estimated using the correlation described by Freeman et al. \cite{freeman2017asmall},

\begin{equation}\label{eq16}
Cost_{HE} = 18.3 \times \dot{W}_{C/E}
\end{equation}

The cost for the pump can be estimated using the correlation described by Freeman et al. \cite{freeman2017asmall},

\begin{equation}\label{eq17}
Cost_{p} = 3048.8 \times \dot{W}_{p}
\end{equation}

\begin{figure}[!t]
    \centering{
  \includegraphics[width=0.7\columnwidth]{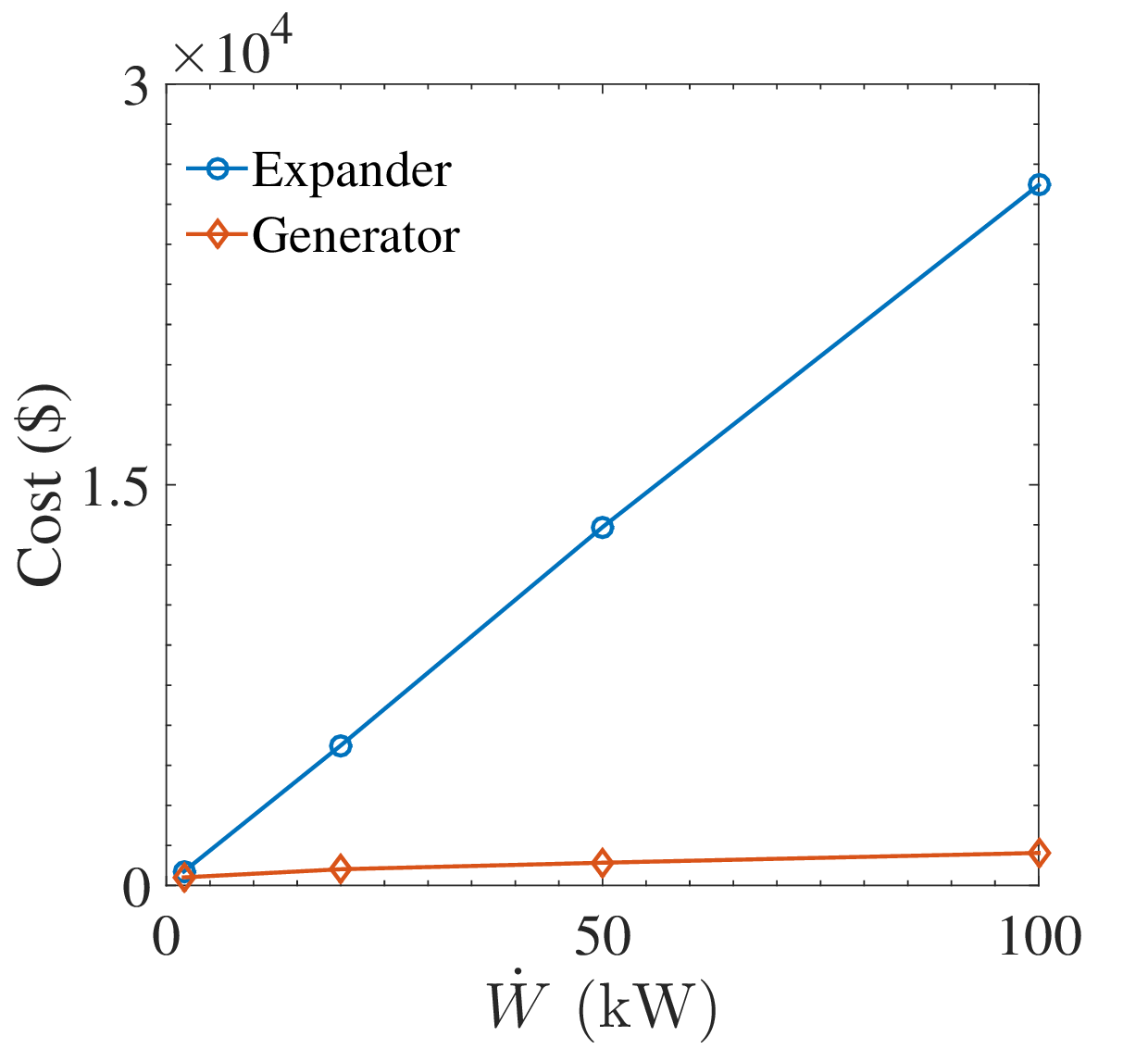}\\ 
  (a)\\
   \includegraphics[width=0.7\columnwidth]{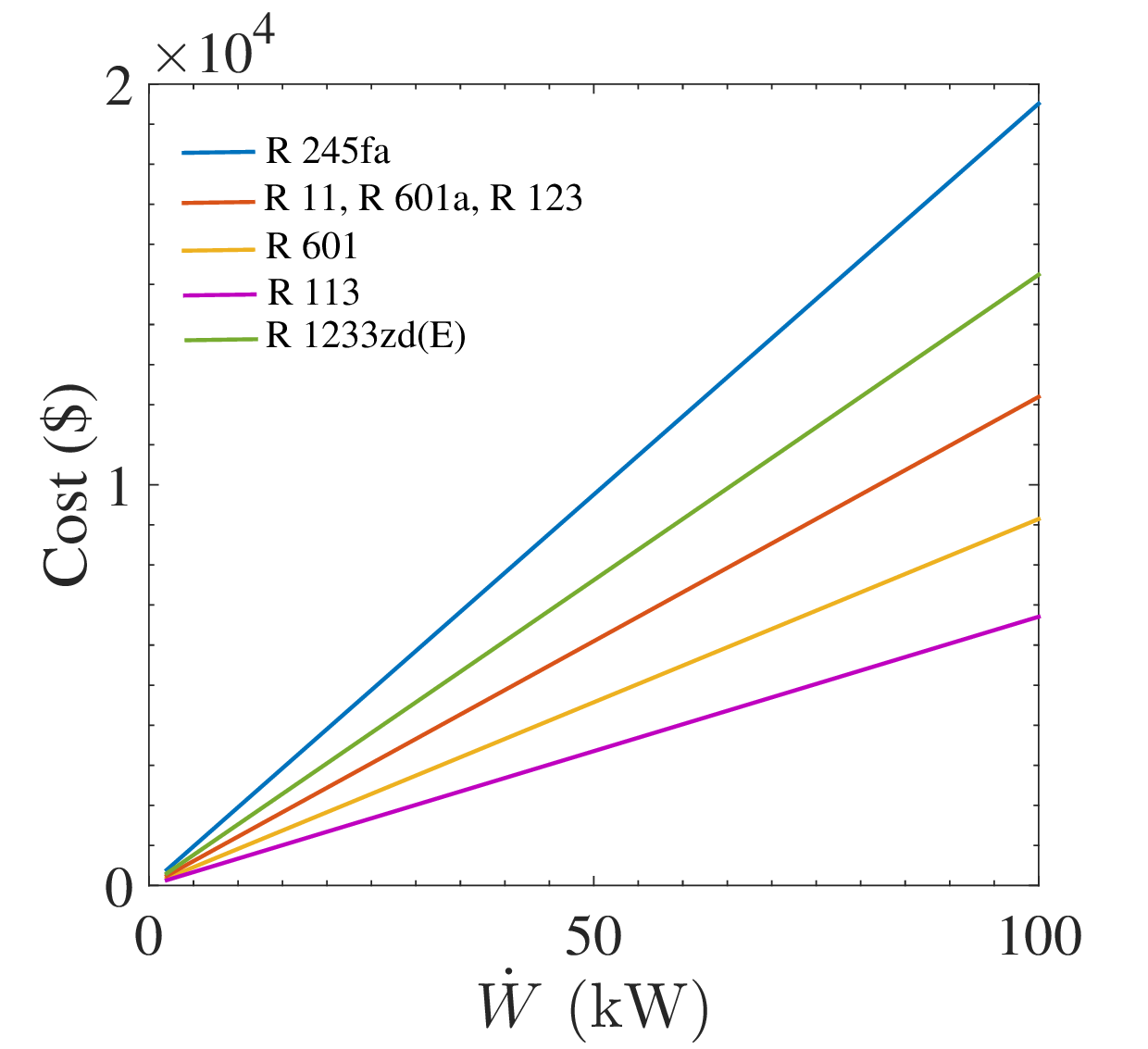}\\
 (b)\\
   \includegraphics[width=0.7\columnwidth]{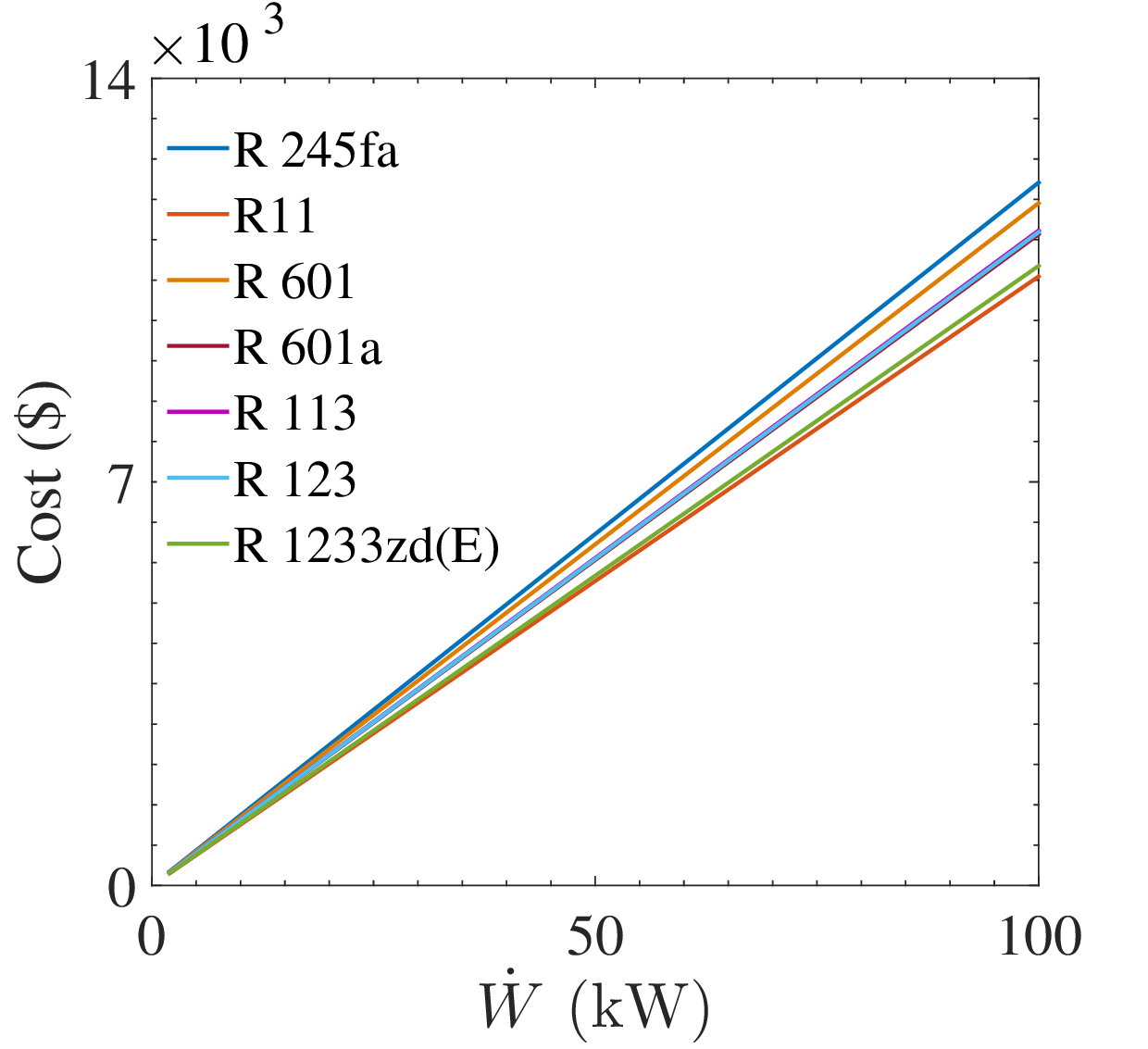}\\
 (c)
    \caption{Cost estimation of Main ORC components\\(a) Expander and generator (b) Pump (c) Heat exchangers}
    \label{cost_main_components}}
\end{figure}

Figure \ref{cost_main_components} describes the cost of the main components of an ORC in terms of power output. The expander contributes the most whereas the generator's contribution is the least.
\mbox{}

\begin{figure}[!t]
    \centering{
  \includegraphics[width=0.8\columnwidth]{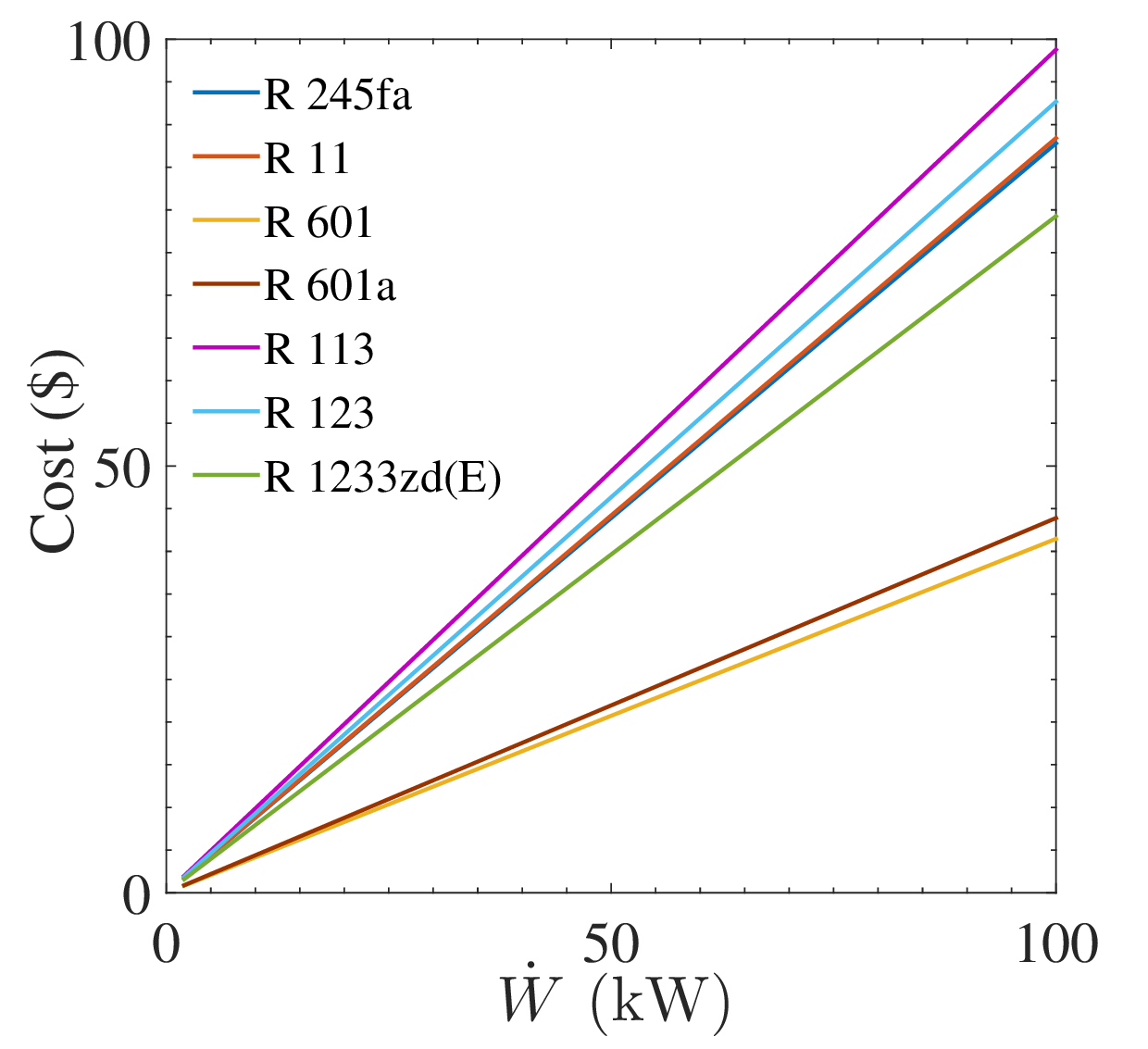}
    \caption{Estimated working fluid cost at different power}
    \label{cost_working_fluid}}
\end{figure}

\begin{figure}[!t]
    \centering{
  \includegraphics[width=0.8\columnwidth]{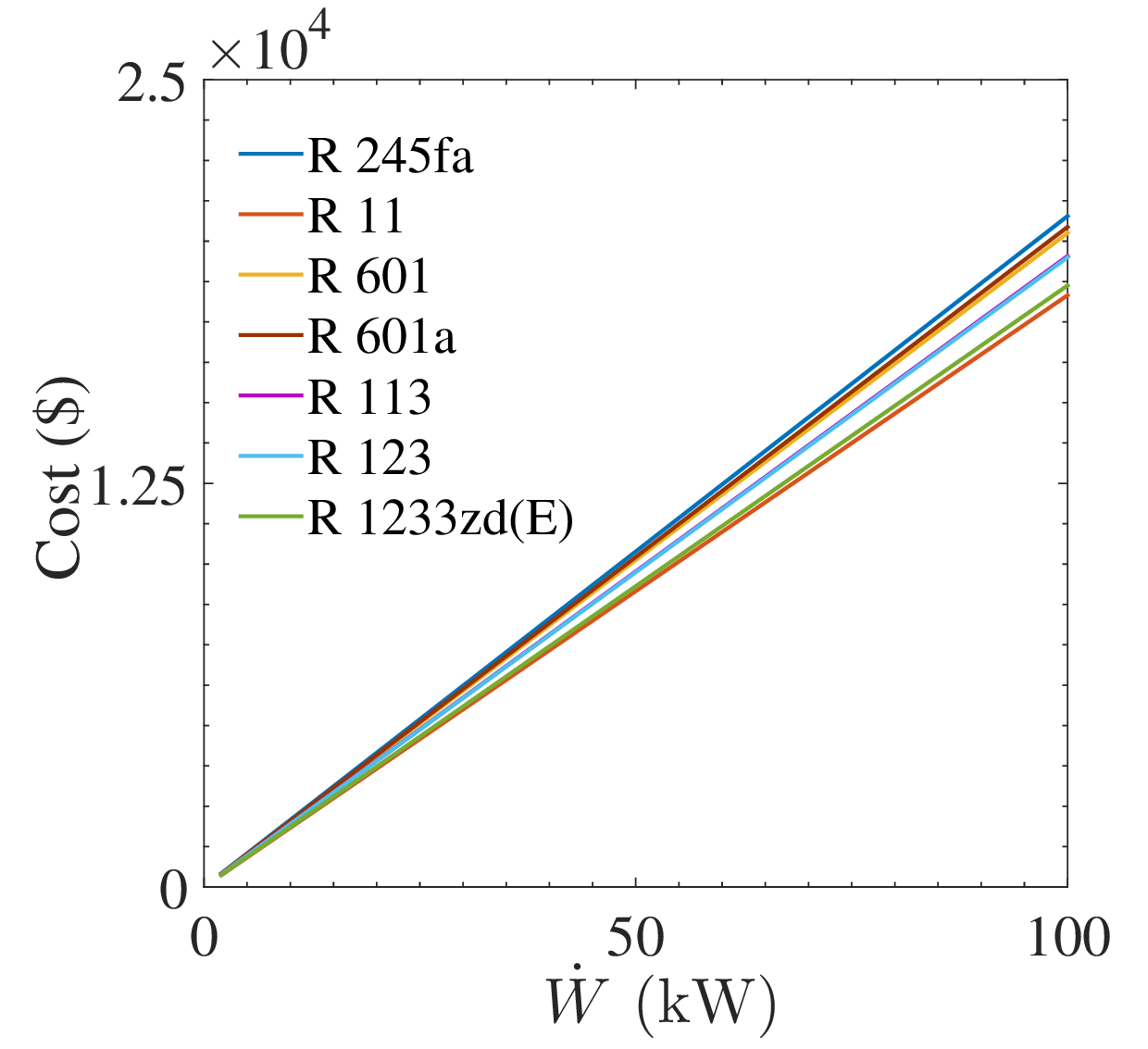}
    \caption{Estimated cost of Solar collector at different power output}
    \label{cost_sc}}
\end{figure}

\begin{figure}[!t]
    \centering{
  \includegraphics[width=0.8\columnwidth]{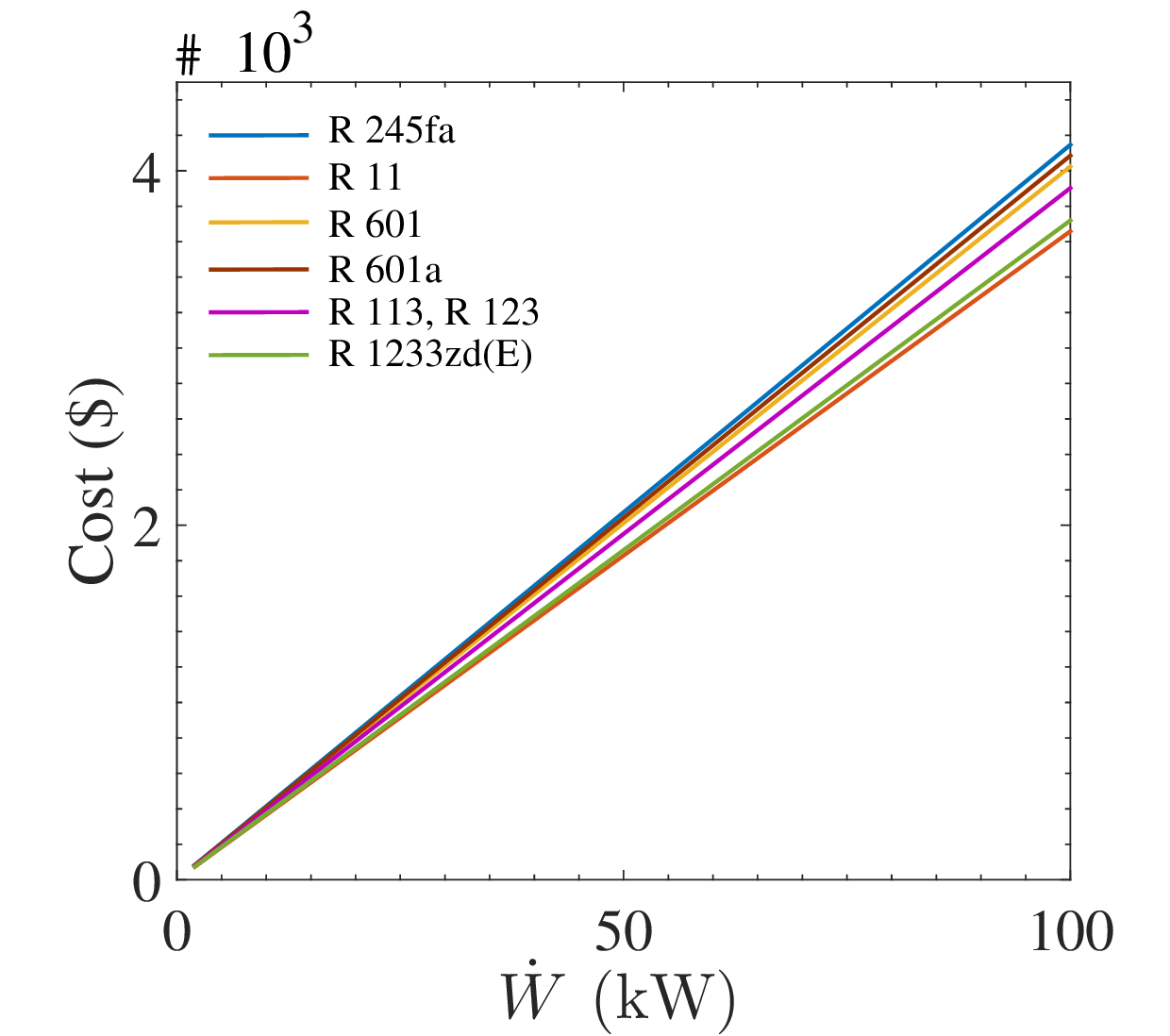}
    \caption{Estimated HTF cost at different power output}
    \label{cost_htf}}
\end{figure}

The cost of the working fluid  can be calculated using the correlation described in \cite{freeman2017asmall},
\begin{equation}\label{eq18}
Cost_{WF} = 30.5 \times \dot{m}
\end{equation}
Figure \ref{cost_working_fluid} describes the cost of the working fluid concerning the power requirement. The lowest cost estimated here for R 601 whereas R 245fa has the maximum cost.
For the estimation of solar collector cost, we need to have an estimate of the required area of the solar collector for targeted power output.
Figure \ref{isoarea} suggests the required collector area required for each scenario for every working fluid. The estimated cost for solar collectors can be expressed as described by Freeman et al. \cite{freeman2017asmall},
\begin{equation}\label{eq20}
    Cost_{col} = 18.3 \times A_{col} 
\end{equation}
Figure \ref{cost_sc} describes the probable cost of a solar collector at a different target power of an ORC. It is observed that the investment cost can be an order of magnitude higher if there is a requirement for a larger collector area. R 11 and R 1233zd(E) require the least investment cost for the solar collector.
The cost for heat transfer fluid can be estimated using the correlation described in \cite{freeman2017asmall}.
\begin{equation}\label{eq21}
  Cost_{htf} = 6.1 \times A_{col} 
\end{equation}
Figure \ref{cost_htf} shows the trend for HTF cost at different target power for different fluids. The estimated cost can be within the range of 2.5-3 lacs.

\begin{figure}[!t]
    \centering{
  \includegraphics[width=0.8\columnwidth]{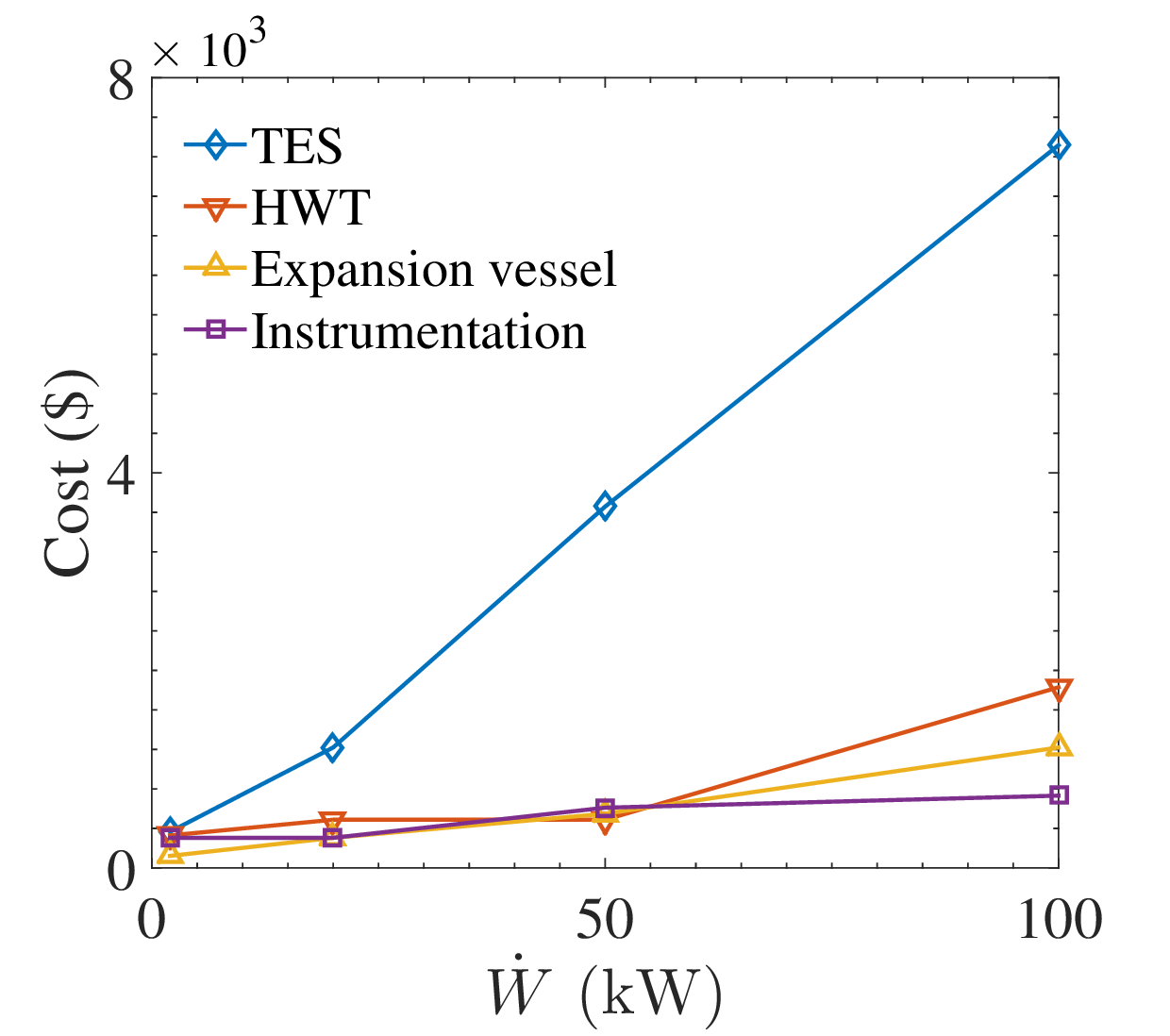}
    \caption{Estimated cost of other ancillary components at different power output}
    \label{cost_other}}
\end{figure}

In the other cost part, thermal energy storage is the component that contributes the most and can reach up to six lacs for the assessment. A hot water tank might not be necessary for applications that don't include the CHP process but might be a good candidate for evaluation in the hilly region where a hot water tank can be beneficial. The cost of the hot water tank, expansion vessel, and thermal storage tank are estimated according to the Indian market scenario.
The figure \ref{cost_other} describes the estimated cost for TES, HWT, Expansion valve, and Instrumentation at different power.

\begin{figure}[!t]
    \centering{
  \includegraphics[width=0.8\columnwidth]{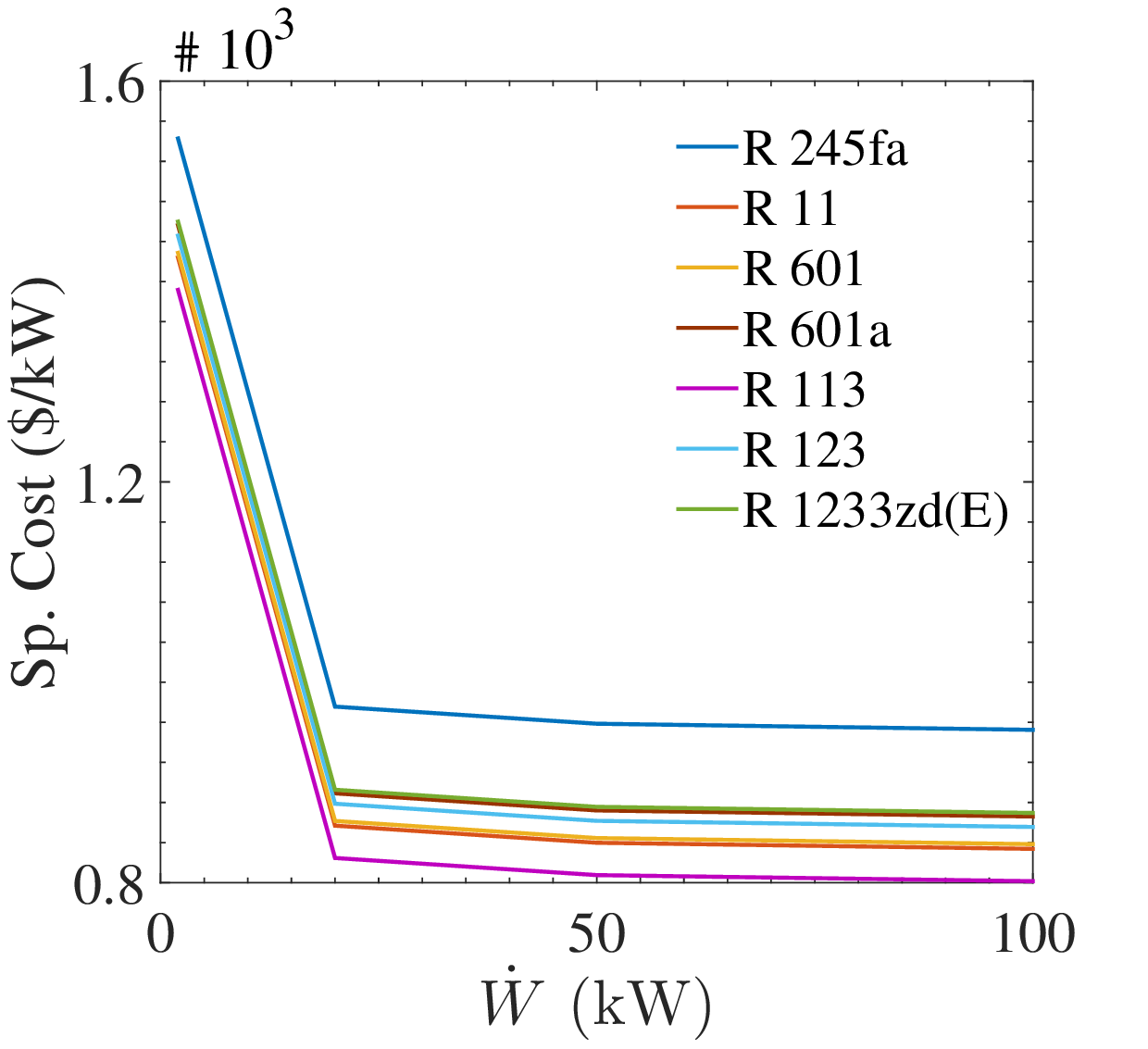}
    \caption{System specific cost at different power output}
    \label{sp_cost}}
\end{figure}

The specific cost can be expressed as the ratio of the net cost to the target power of the system. For our assessment, the specific cost of the product is lowest for R 113 or R 11 and highest for R 245fa. Figure \ref{sp_cost} shows the trend of specific costs for all the working fluids.

\subsection{Optimal performing fluid}
After evaluating the cost and system efficiency, we choose the optimum working fluid for s-ORC following a grading approach. We ranked each fluid from 1 to 7 for three different categories- 1. Efficiency, 2. Specific cost, 3. Environment friendliness. The marks for each rank follow the reverse order of 7 to 1 (7 marks and 1 mark for $1^{st}$ and $7^{th}$ ranked working fluid). Upon giving equal weightage to all three indicators, we present the total grade of each working fluid candidate in table \ref{optimum_fluid}. Using a similar procedure, we further rank the working fluids for other scenarios presented in table \ref{optimum_fluid_other}.

\begin{table}[htb]
    \centering
        \caption{Grading table for the working fluids for scenario 1}
    \label{optimum_fluid}
    \begin{tabular}{c c c c c c}
    \hline
     WF  &  $\eta$ & SC & Env. Impact & Total & Rank \\
     \hline
    R 245fa & 1 & 1 & 3 & 5 & 7\\
    R 11 & 7 & 6 & 1 & 14 & 2\\
    R 601 & 3 & 2 & 5.5 & 10.5 & 5\\
    R 601a & 2 & 2 & 5.5 & 9.5 & 6\\
    R 113 & 4 & 7 & 2 & 13 & 3\\
    R 123 & 5 & 3 & 4 & 12 & 4\\
    R 1233zd(E) & 6 & 4 & 7 & 17 & 1 \\
    \hline
    \end{tabular}
\end{table}
Notably, two working fluids- R 601a and R 601 have ranked $2^{nd}$ in the environment friendliness category. Hence, each of them is graded an average of the total grade for $2^{nd}$ and $3^{rd}$ rank in that respective category. The table \ref{optimum_fluid} and table \ref{optimum_fluid_other} suggest that R 1233zd(E) is the optimum performing fluid at the conditions considered in the present work.

\begin{table}[htb]
    \centering
        \caption{Rank of working fluids in other scenarios}
    \label{optimum_fluid_other}
    \begin{tabular}{|c|ccc|}
    \hline
        \diagbox{WF}{Rank} & scenario 2 & scenario 3 & scenario 4 \\
        \hline
        R 245fa & 7& 6 & 6 \\
        R 11 & 2 & 4& 1\\
        R 601 & 5 & 3& 3\\
        R 601a & 4 &5 & 5\\
        R 113 & 1 & 1& 4 \\
        R 123 & 6 & 7& 7\\
        R 1233zd(E) &2 & 1& 1\\
        \hline
    \end{tabular}
\end{table}
\subsection{Comparison of cost with the existing technologies}\label{cost_comparison_v0p1}

The initial scenario-based evaluation gives us an estimation and comparison of cost while using different working fluids. Upon shortlisting the R 1233zd(E) as a choice of the selected candidate, we compare the estimated specific investment cost with technologies existing for other renewable technologies available. The table \ref{cost_comparison} summarizes the cost comparison at 20,50 and 100 kW of targeted power for each scenario. The cost for S-ORC indicates that this technology can be competitive with the existing technologies available in the market.

\begin{table}[!h]
    \centering
        \caption{Comparison of cost reported in the existing technology with S-ORC}
    \label{cost_comparison}
    \begin{tabular}{ |c | c |c |c|}
    \hline
    Targeted power & 20 kW & 50 kW & 100 kW \\
     \hline
      Technology & \multicolumn{3}{c|}{Specific cost in USD/kWh}\\
      \hline
      Solar PV & 1148.78 & 1042.683 & 987.8049\\
      Biomass-based ORC & 3658.537 & 3048.78 & 2817.073\\
      S-ORC scenario 1 & 918.1463 & 894.9634 & 894.9634\\
      S-ORC scenario 2 & 966.4268 & 943.2317 & 943.2317\\
      S-ORC scenario 3 & 1016.11 & 992.939 & 992.939\\
      S-ORC scenario 4 & 1145.268 & 1122.098 & 1122.098\\
      \hline
    \end{tabular}
\end{table}

\section{Conclusion and further scope}\label{conclusion}
The work aims to identify and assess the viability of a solar-powered Organic Rankine cycle unit in the Indian context. A traditional ORC system is considered where the source and sink temperatures are 363 K-423 K and 298K, respectively. Using this condition, we summarize the list of objectives which are as follows:
\begin{itemize}
\item Identifying several suitable organic fluids based on several indicating factors mentioned in section 3.1.
\item Parametric evaluation of each working fluid to identify the operational envelope.
\item To assess the cycle's performance for each working fluid using a four-scenario approach.
\item Estimate the cost of the system for all working fluids.
\item Determine the best working fluid and compare the cost of the system to existing technologies.
\end{itemize}
The key findings in this article are summarized below.
\begin{itemize}
\item System using R-11 as working fluid offers the best cycle efficiency in all scenarios. But, working fluids offering the lowest efficiency are different in all scenarios.
\item The least efficient working fluid is R 245fa in scenario 1, followed by R 601 in scenario 2, and R 123 in scenarios 3 and 4.
\item For any scenario, R-11-containing systems require the highest flow rate and N-pentane (R-601) containing systems require the lowest flow rate.
\item For scenarios 1 and 2, systems using R 245fa as the working fluid require the largest collector area, whereas, for scenarios 3 and 4, the collector area requirement is the largest for R 123. R-11, however, gives the minimum collector area requirement for each scenario.
\item R 113 is the best working fluid candidate in minimizing the system cost for each scenario.
\item Considering efficiency, system cost, and environmental friendliness as indicators with equal weightage, we identify R 1233zd(E) as the optimum working fluid candidate.
\item Considering three indicators, the cost of the system having R 1233zd(E) as the working fluid shows promising results concerning techno and economic viability compared to the preexisting technologies.
\end{itemize}
The S-ORC considered for analysis is the simplest version of ORC. We envisage that the achievable performance could be better by using special techniques like preheating; furthermore, the cost of the system majorly depends on the solar collector and the expander system. Selecting and optimizing the expander and solar collector suitable for such applications could reduce the cost and improve conversion efficiency, thus leaving a plausible scope for further research.
\bibliography{reference}
\end{document}